\documentclass[12pt]{aastex}

\usepackage{graphicx,natbib,textcomp,caption,subcaption}         
\usepackage[utf8]{inputenc}
\usepackage{xcolor,amsmath}
\usepackage{multirow}
\usepackage{float}
\usepackage{hyperref}
\usepackage{xfrac}

\slugcomment{ApJ manuscript, \today}

\begin{document}

\title{Three-dimensional simulations of the inhomogeneous Low Solar Wind}

\author{N. Magyar, V. M. Nakariakov}
\affil{Centre for Fusion, Space and Astrophysics, Physics Department, University of Warwick, Coventry CV4 7AL, UK; norbert.magyar@warwick.ac.uk}

\begin{abstract}
In the near future, \textit{Parker Solar Probe} will put theories about the dynamics and nature of the transition between the solar corona and the solar wind to stringent tests. The most popular mechanism aimed to explain the dynamics of the nascent solar wind, including its heating and acceleration is magnetohydrodynamic (MHD) turbulence. Most of the previous models focus on nonlinear cascade induced by interactions of outgoing Alfv\'en waves and their reflections, ignoring effects that might be related to perpendicular structuring of the solar coronal plasma, despite overwhelming evidence for it. In this paper, for the first time, we analyse through 3D MHD numerical simulations the dynamics of the perpendicularly structured solar corona and solar wind, from the low corona to $15\ \mathrm{R}_\odot$. We find that background structuring has a strong effect on the evolution of MHD turbulence, on much faster time scales than in the perpendicularly homogeneous case. On time scales shorter than nonlinear times, linear effects related to phase mixing result in a 1/f perpendicular energy spectrum. As the turbulent cascade develops, we observe a perpendicular (parallel) energy spectrum with the power law index of -3/2 or -5/3 (-2), a steeper perpendicular magnetic field than velocity spectrum, and a strong build-up of negative residual energy. We conclude that the turbulence is most probably generated by the self-cascade of the driven transverse kink waves, referred to previously as `uniturbulence', which might represent the dominant nonlinear energy cascade channel in the pristine solar wind.           
\end{abstract}

\section{Introduction}

Predicted by \cite{1958ApJ...128..664P} and detected in-situ soon after \citep{1960SPhD....5..361G,1962Sci...138.1095N}, the solar wind is a continuous outflow of charged particles originating from the Sun. Our understanding of the dynamics and nature of the solar wind has come a long way since \citep{1995SSRv...73....1T,2013LRSP...10....2B,2019LRSP...16....5V}, however there are still many questions unanswered. These concern the source of energy for the acceleration of the solar wind, the structure and dynamics of the near-Sun solar wind, and the acceleration and transport mechanism of energetic particles, among others \citep{2016SSRv..204....7F}.  \par
The solar wind is generally thought to be driven by magnetohydrodynamic (MHD) turbulence, especially the fast wind in open field areas. Current MHD turbulence driven solar wind models differ, among other aspects, in their underlying turbulence generation mechanism. It is understood that nonlinear interactions of fluctuations, leading to nonlinear cascade, drive turbulence. However, the nature of the fluctuations and their nonlinear interactions depend on the specific models considered. The first phenomenological model of MHD turbulence was that of an incompressible and homogeneous plasma, in which the fluctuations are Alfv\'en waves propagating parallel and anti-parallel to some mean magnetic field. Nonlinear interactions occur when oppositely propagating Alfv\'en waves collide, resulting in a nonlinear cascade which can be thought of as a mutual deformation of the Alfv\'en waves \citep{1964SvA.....7..566I,1965PhFl....8.1385K,2013PhPl...20g2302H}. To date, most of the modelling relies on the counterpropagating Alfv\'en wave phenomenology. In these models, waves propagating outward from the Sun are thought to originate in the lower corona, and ultimately in the convective hydrodynamic buffeting of the photosphere. There is observational evidence that outward-propagating transverse waves are omnipresent in the solar corona \citep[e.g.][]{2007Sci...317.1192T,2007Sci...318.1574D}. In open field regions, sunward propagating waves are thought to owe their existence to non-WKB reflection, on the Alfvén speed gradients along the propagation direction \citep{1980JGR....85.1311H,2007JGRA..112.8102H}. Then, MHD turbulence is generated by the nonlinear interaction of outward propagating and reflected Alfv\'en waves, sometimes referred to as Alfv\'en wave turbulence. Analytical and numerical solar wind models with Alfv\'en wave turbulence usually rely on a reduced MHD treatment \citep[e.g.,][]{1999ApJ...523L..93M,2001PhPl....8.2377D,2002ApJ...575..571D,2011ApJ...736....3V,2013ApJ...776..124P,2014ApJ...787..160W,2016ApJ...821..106V,2017ApJ...835...10V,2019JPlPh..85d9009C}, or on one-dimensional Alfvén wave equations with approximate terms for reflection and turbulent dissipation or Reynolds-averaging, as encountered also in global solar wind models \citep[e.g.,][, although note the 3D incompressible MHD global solar wind model of \citealt{2017ApJ...837...75S}]{2005ApJS..156..265C,2014ApJ...796..111L,2014ApJ...788...43U,2014ApJ...782...81V,2018ApJ...865...25U}, and even on fully compressible MHD \citep{2012ApJ...749....8M,2014MNRAS.440..971M,2018ApJ...853..190S,2019ApJ...880L...2S}, although the compressible MHD simulations usually contain also other nonlinear cascade channels besides the counterpropagating wave scenario, such as wave steepening or shock formation, among others. In some of these solar wind models, it is found that heating of the open magnetic field regions by Alfv\'en wave turbulence alone is insufficient \citep{2010ApJ...708L.116V,2016ApJ...821..106V,2017ApJ...835...10V,2019SoPh..294...65V}, hinting that some additional mechanism besides the reflection-driven Alfv\'en wave turbulence is missing. \par 
Moving beyond the countepropagating Alfv\'en wave phenomenology, other models consider a richer spectrum of fluctuations and more channels for nonlinear interactions. In the Nearly-Incompressible (NI) MHD formalism, the fluctuations are divided into incompressible, non-propagating quasi two-dimensional (2D) and propagating NI `slab' fluctuations \citep[e.g.,][]{1993PhFlA...5..257Z,2017ApJ...835..147Z}, justified by the appearance of  solar wind fluctuations as a superposition of dominant 2D and minority slab components \citep[e.g.,][and references therein]{1990JGR....9520673M,2013LRSP...10....2B}. The 2D and slab fluctuations are transverse to the mean magnetic field, but either vary only across the field (i.e. $k_\parallel = 0$), or along it (i.e. $k_\perp = 0$), respectively. An example of 2D fluctuations are `vortex' magnetic field solutions \citep{2017ApJ...835..147Z}, while Alfv\'en waves with $k_\perp = 0$ are an example of incompressible slab fluctuations. In the NI MHD studies, it is shown that nonlinear interactions take place both in and between the 2D and slab components. For example, the nonlinear evolution of the incompressible 2D component is not mediated by waves, but proceeds in an essentially 2D MHD fashion orthogonal to the mean magnetic field. Studies based on the NI MHD formalism show that the nonlinearity of the 2D or quasi-2D component is the dominant nonlinear cascade channel in the solar corona and solar wind \citep{2017ApJ...851..117A,2018ApJ...854...32Z,2020ApJ...901..102A}. Moreover, unidirectionally propagating Alfv\'en waves can interact nonlinearly with the 2D components, resulting in their cascade towards a Kolmogorov $-5/3$ power spectrum \citep{2019ApJ...887..160T,2020ApJ...898..113Z,2020ApJ...900..115Z}. \par 
It is probably noteworthy here to reinterpret the counterpropagating Alfv\'en wave scenario in the light of the NI MHD formalism. For nonlinear interactions to occur between colliding Alfv\'en waves, their wavefronts must mutually vary in the perpendicular directions, expressed as $\mathbf{k}_\perp^- \times \mathbf{k}_\perp^+ \neq 0$ \citep[e.g.,][]{2013PhPl...20g2302H}. In this sense, the interacting Alfv\'en waves can be regarded as displaying both 2D and slab components, and their interactions as already inclusive of mixed 2D and slab nonlinear interactions. \par 
Despite the competing ideas for the dominant nonlinear cascade channel, a universal feature of all the previous non-global solar wind modeling efforts is homogeneity of the plasma perpendicular to the mean (radial) magnetic field. This is a strikingly obvious deficiency of the current models when considering that there is ample evidence of structuring \citep[e.g.][]{2014ApJ...788..152R,2016JGRA..121.5055B,2016ApJ...828...66D,2018ApJ...860...34H,2020ApJ...893...64G,2020ApJS..246...60P,2020ApJS..246...57K,10.3389/fspas.2020.00020}, which ``belie the notion of a smooth outer corona" \citep{2018ApJ...862...18D}, with variations in density up to an order of magnitude on spatial scales of 50 Mm. It is important to point out here that by the structuring of the plasma we mean perpendicular inhomogeneities of the background plasma density, radial flow speed, radial magnetic field, etc., which lead effectively to perpendicular Alfv\'en speed gradients, and not the perpendicular inhomogeneity of fluctuations or perturbations in velocity, magnetic field, etc. In fact, perpendicular structuring of fluctuations, that is, $k_\perp \neq 0$ is an essential prerequisite of all nonlinear interactions which cascade energy perpendicularly, including that of counterpropagating Alfv\'en wave interactions, as noted above. Therefore, all of the solar wind models cited above rely on the perpendicular structuring of the fluctuations. However, the structuring of the background plasma is not included self-consistently in these works. Instead, it is represented by a characteristic perpendicular lengthscale or correlation length of the fluctuations, which is the case even in global solar wind models with inhomogeneous background. Even though plasma structuring is not included in existing non-global solar wind models initially, it is worth noting that in some compressible MHD models structuring develops from the ensuing turbulent dynamics \citep{2018ApJ...859L..17S,2019ApJ...880L...2S}. \par
It is already clear from the discussion above that the nonlinear interaction of counterpropagating Alfv\'en waves is not the only channel for turbulence generation in MHD. However, if furthermore transverse background inhomogeneities are included, the spectrum of wave solutions admitted by the MHD equations \citep{2011SSRv..158..289G,10.3389/fspas.2019.00020} and their nonlinear interactions becomes much richer, already in the incompressible case \citep{1989JPlPh..41..479M,1990PhRvL..64.2591Z,2019ApJ...882...50M}. NI MHD equations with inhomogeneous background were also derived \citep{2010ApJ...718..148H,2017ApJ...835..147Z}. If one considers the fully compressible MHD equations, the spectrum is even richer \citep{1987JGR....92.7363M,2013PhRvE..87a3019B,2016ApJ...829L..27B,2018JPlPh..84d9004A}. \citet{2019ApJ...882...50M} showed that inhomogeneities perpendicular to the magnetic field admit transverse wave solutions with different properties when compared to those of pure Alfv\'en waves, with important nonlinear implications. Some of these transverse waves, referred to mostly as kink\footnote{In some works of other authors, and in some of our previous works \citep{2017NatSR...14820M,2019ApJ...873...56M,2019ApJ...882...50M} these waves are referred to as Alfv\'enic instead, following \citet{2009A&A...503..213G}, to reflect the fact that they are driven mostly by magnetic tension, as pure Alfv\'en waves. In this work, we prefer to use the term `kink', in order to better emphasize the differences between these waves and pure Alfv\'en waves.} waves are propagating on structures along the magnetic field (e.g., on a magnetic flux tube of higher density than the surrounding plasma), manifest as propagating transverse displacements of these structures, and can self-deform or self-cascade nonlinearly, without the need for counterpropagating waves or other fluctuations being present. To see why this is the case, \citet{2019ApJ...882...50M} employed the Els\"{a}sser formulation of the MHD equations. \par 
Els\"asser variables \citep{1950PhRv...79..183E} are usually employed in solar wind studies to separate the outward propagating waves (denoted by $z^+$) and the reflected or inward propagating waves (denoted by $z^-$). The separation is exact for even fully nonlinear, unidirectionally propagating waves in homogeneous and incompressible plasma, i.e. Alfv\'en waves, and it even holds for radially inhomogeneous (along a purely radial magnetic field) but otherwise homogeneous plasma without nonlinear interactions \citep{2007JGRA..112.8102H,2019ApJ...882...50M}. 
However, beyond pure Alfv\'en wave dynamics, it is often overlooked that transverse inhomogeneities, compressibility, and the nonlinear interaction of waves renders the separation of fluctuations into inward and outward propagating waves inexact. For example, inhomogeneity and compressibility allows for waves (e.g. fast, slow MHD waves, surface Alfv\'en waves, kink waves, etc.) that are mostly described by both Els\"asser variables as they propagate \citep{2019ApJ...873...56M,2019ApJ...882...50M}. In fact, waves other than pure Alfv\'en waves generally perturb both Els\"{a}sser variables as they propagate. While kink waves, both propagating and standing, are routinely observed in the corona \citep[e.g.,][]{1999Sci...285..862N,2007Sci...317.1192T,2015A&A...583A.136A,2016GMS...216..395W,2019ApJS..241...31N}, evidence of surface Alfv\'en waves in the solar wind is as of yet inconclusive \citep[e.g.,][]{2001GeoRL..28..677H,2001JGR...10629283V,2013AnGeo..31..871P}. 
Besides waves that are not pure Alfv\'en waves, structures (inhomogeneities) advected by the solar wind also perturb both Els\"asser variables \citep{1990JGR....95.4337T,1995SSRv...73....1T,2012ApJ...745...35Z,2015ApJ...805...63A}. The nonlinear interaction of Alfv\'en waves can generate purely magnetic fluctuations, 2D modes ($k_\parallel = 0$) or condensates which as well perturb both Els\"asser variables \citep{2009PhRvL.103v5001B,2013PhPl...20g2302H}. Indeed, the nature of the inward $z^-$ Els\"asser variable is often not clear \citep{2018JGRA..123...57W}.
Previous studies on Alfv\'en wave dynamics in radially inhomogeneous models often mention the existence of an `anomalous' $z^-$ component which is co-propagating with $z^+$ \citep{1989PhRvL..63.1807V,2009ApJ...700L..39V,2013ApJ...776..124P}. The issue of `anomalous' waves is solved by \citet{2007JGRA..112.8102H}, who shows that, while the continuously generated, reflected $z^-$ components might show up as co-propagating in a harmonic analysis, their impulse response analysis shows that these reflected Alfv\'en waves still follow sunward characteristics, i.e. that there are no truly co-propagating Els\"{a}sser variables in these studies. Nevertheless, the coherence of the Els\"{a}sser variables resulting from this linear coupling of Alfv\'en waves seems to influence their spectrum \citep{2009ApJ...700L..39V}. \par
The key observation coming from the Els\"{a}sser formulation of the incompressible MHD equations is that nonlinear advective terms responsible for the nonlinear interactions of, among others, counterpropagating Alfv\'en waves, require both Els\"{a}sser fields to be nonzero. As discussed above, for waves in a homogeneous incompressible plasma, the two Els\"{a}sser variables represent Alfv\'en waves propagating parallel and anti-parallel to the mean magnetic field, so the nonlinear condition implies the presence of interacting, oppositely propagating Alfv\'en waves. However, in the case of e.g. kink waves, the same nonlinear condition is already satisfied by a unidirectionally propagating wave, as kink waves are described by co-propagating Els\"{a}sser fields \citep[see Eq. 21 in ][]{2019ApJ...882...50M}. Note that surface Alfv\'en and kink wave solutions are already admitted in incompressible MHD with magnetic field and background flow inhomogeneities, but homogeneous background density. Under these conditions, the essential nonlinear term responsible for the nonlinear cascade is still the nonlinear advective term requiring both Els\"{a}sser fields nonzero. However, allowing for background density inhomogeneities and compressibility introduces more nonlinear terms of different forms (see citations in the discussion above), which do not necessarily require both Els\"{a}sser variables to be nonzero, and which might each contribute in part to the generation of turbulence. \par
The \textit{Parker Solar Probe} (\textit{PSP}) offers an unprecedented experimental insight and test to our understanding of the pristine solar wind. Therefore, constructing adequate models which will be compared to \textit{PSP} observations is crucial and very timely. This study aims to fill in the considerable gap in our understanding of wave propagation and dynamics in the intrinsically inhomogeneous solar corona and nascent solar wind. We focus on the dynamics rather than the acceleration and heating of the solar corona and wind in this paper. The paper is structured as follows: in Section~\ref{two}, the numerical method and model are presented, in Section~\ref{three}, we present the results of the simulations, along with discussions, and in Section~\ref{four} we conclude the findings. 

\section{Numerical model and method}
\label{two}

\begin{figure}[h]
    \centering     
        \includegraphics[width=1.0\textwidth]{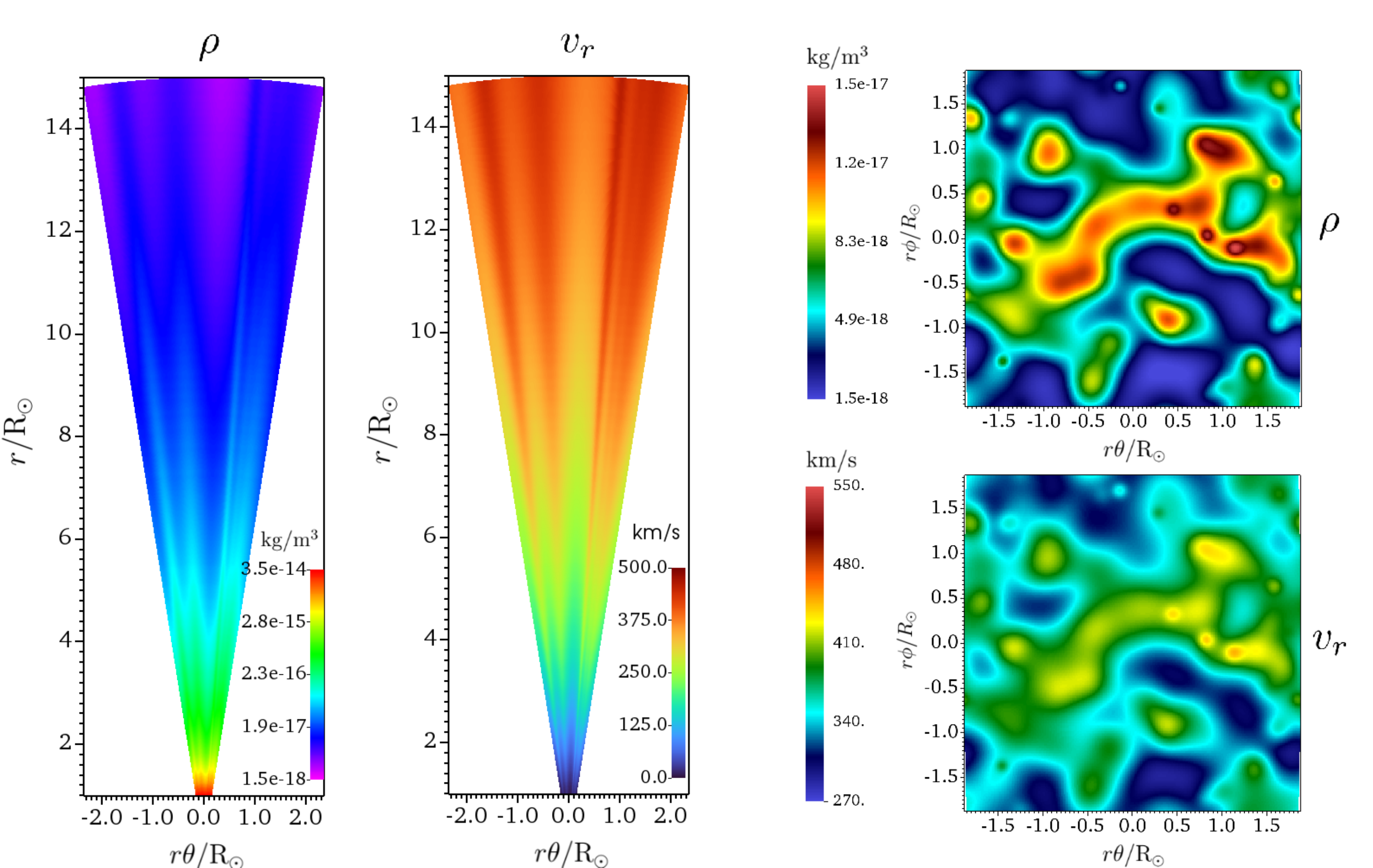}  
        \caption{Snapshots of the initial condition, showing the density ($\rho$) and radial velocity ($v_r$) in a slice along $\phi = \pi/2$ (left), and in a spherical slice at $r = 12\ R_\odot$. (right)}
        \label{initcond}
\end{figure}

We run full 3D, compressible and ideal MHD simulations using \texttt{MPI-AMRVAC} \citep{2014ApJS..214....4P,2018ApJS..234...30X}, using the three-step, third-order \texttt{HLLD} solver with \texttt{vanLeer} second-order slope limiter. The solenoidality of the magnetic field is maintained by using the constrained transport method.  
A slightly modified spherical geometry is implemented, in which for a numerical domain centred at $\theta=\pi/2$ the $\theta$ and $\phi$ angular directions are symmetrized by removing the $\mathrm{sin(\theta)}$ dependence of $d \phi$, as employed in e.g.,  \citet{2019ApJ...880L...2S}. This is enabling the use of periodic boundary conditions in both $\theta$ and $\phi$ directions. The numerical domain extends from $1.01\ R_\odot$ to $15\ R_\odot$, and $0.1\pi\ \mathrm{rad}$ in the angular directions.  Therefore, the radial bottom boundary corresponds to the low corona ($\approx  7\ \mathrm{Mm}$ from the photosphere). Gravity is included using the inverse-square law from the solar surface. As we intend to simulate an open field region of the solar corona and solar wind, the initial magnetic field is purely radial, with a flux density of 2 G at the bottom boundary. The super-radial expansion of the magnetic field close to the solar surface is neglected \citep[e.g.][]{1976SoPh...49...43K}. A solar wind solution is prescribed initially by interpolating tabulated, parametrized 1D wind solutions onto the 3D grid. The 1D solutions are obtained by running 1D simulations from $1.01\ R_\odot$ to $15\ R_\odot$ for a range of temperatures at the bottom radial boundary, from $1\ \mathrm{MK}$ to $2.25\ \mathrm{MK}$, while keeping the bottom density constant ($\rho = 3.24 \times 10^{-14}\ \mathrm{kg\ m^{-3}}$). The resulting solar wind solutions, i.e. the values and radial dependence of density, radial velocity, etc., are in good agreement with those found in other studies \citep[e.g.,][]{2005ApJS..156..265C,2019JPlPh..85d9009C}. For the inhomogeneous 3D solar wind, we impose a temperature map at the bottom radial boundary, and then populate the simulation domain using the 1D solutions. The temperature map is obtained by adding random Gaussian perturbations to a constant temperature background: 
 \begin{equation}
  \begin{array}{lr}
   T_b(r=1.01\ R_\odot,\theta,\phi) = T_0 + \sum\limits_{i=1}^{N} A_{i} \exp^{-[(\theta-\theta_{i})^2 + (\phi-\phi_{i})^2]/ \sigma_{i}^2},
  \end{array}
 \label{Gaussiantemp}
\end{equation}
where $T_0 = 1\ \mathrm{MK}$ is the background temperature, N=200 is the number of Gaussian density enhancements added, $\theta_i, \phi_i, A_i, \sigma_i$ are the random (from uniform distributions) $\theta,\phi$ position, amplitude, and width, respectively, of the $i$\textsuperscript{th} enhancement. The limits of the uniform distributions are the box size $[0.45\pi\ \mathrm{rad},0.55\pi\ \mathrm{rad}]$ for $\theta_i$ and $\phi_i$, $[0,0.25\ \mathrm{MK}]$ for $A_i$, and $[0,0.012\pi\ \mathrm{rad}]$ for $\sigma_i$. The periodicity of the temperature map is assured by wrapping the domain around and therefore allowing the enhancements to pass through the `boundaries'. Although the density is uniform at the bottom, the inhomogeneous temperature map leads to an inhomogeneous solar wind solution higher up in the simulation domain, resulting in both density and flow speed gradients transverse to the radial direction. These random inhomogeneities aim to represent the observed highly structured nature of the solar corona and low solar wind, as described in the Introduction. See Fig.~\ref{initcond} for snapshots of the initial condition. 
Additionally, we run a homogeneous setup with a uniform temperature map $T_b = 1.5\ T_0$, in order to perform a number of comparisons between the homogeneous and inhomogeneous cases. 
In the simulation domain, a space-dependent continuously-operating  heating function is used (adapted from the \texttt{solar wind} test problem included in \texttt{MPI-AMRVAC}), of the form:
 \begin{equation}
  \begin{array}{lr}
   H(r,\theta,\phi) = H_0\rho(r,\theta,\phi)\mathrm{max}\left(T_b(\theta,\phi) - T(r,\theta,\phi),0\right)\mathrm{exp}\left(-((r-1.01\ R_\odot)/H_{sc})^2\right),
  \end{array}
 \label{GaussianHeat}
\end{equation}
where $H_0 = 1.4\ \mathrm{erg/cm^3}$, $\mathrm{max}()$ returns the item with the highest value, and $H_{sc} = 4.5 R_\odot$ is the heating scale height. Employing thermal conduction from the fixed-temperature radial boundary instead of the heating function in Eq.~\ref{Gaussiantemp} results in similar wind solutions, but at a higher computational cost. As the heating and the acceleration of the solar wind is not the main focus of this paper, we opted for the heating function to reduce computational costs. The resulting 3D inhomogeneous wind is not exactly in total pressure equilibrium in the angular directions, and once the simulations starts slight adjustments happen until an equilibrium is reached. The flows resulting from this equilibration are negligible compared to the velocity perturbations induced by the driver, therefore we do not let the system relax before starting the wave driver. \par
\begin{figure}[h]
    \centering     
        \includegraphics[width=0.5\textwidth]{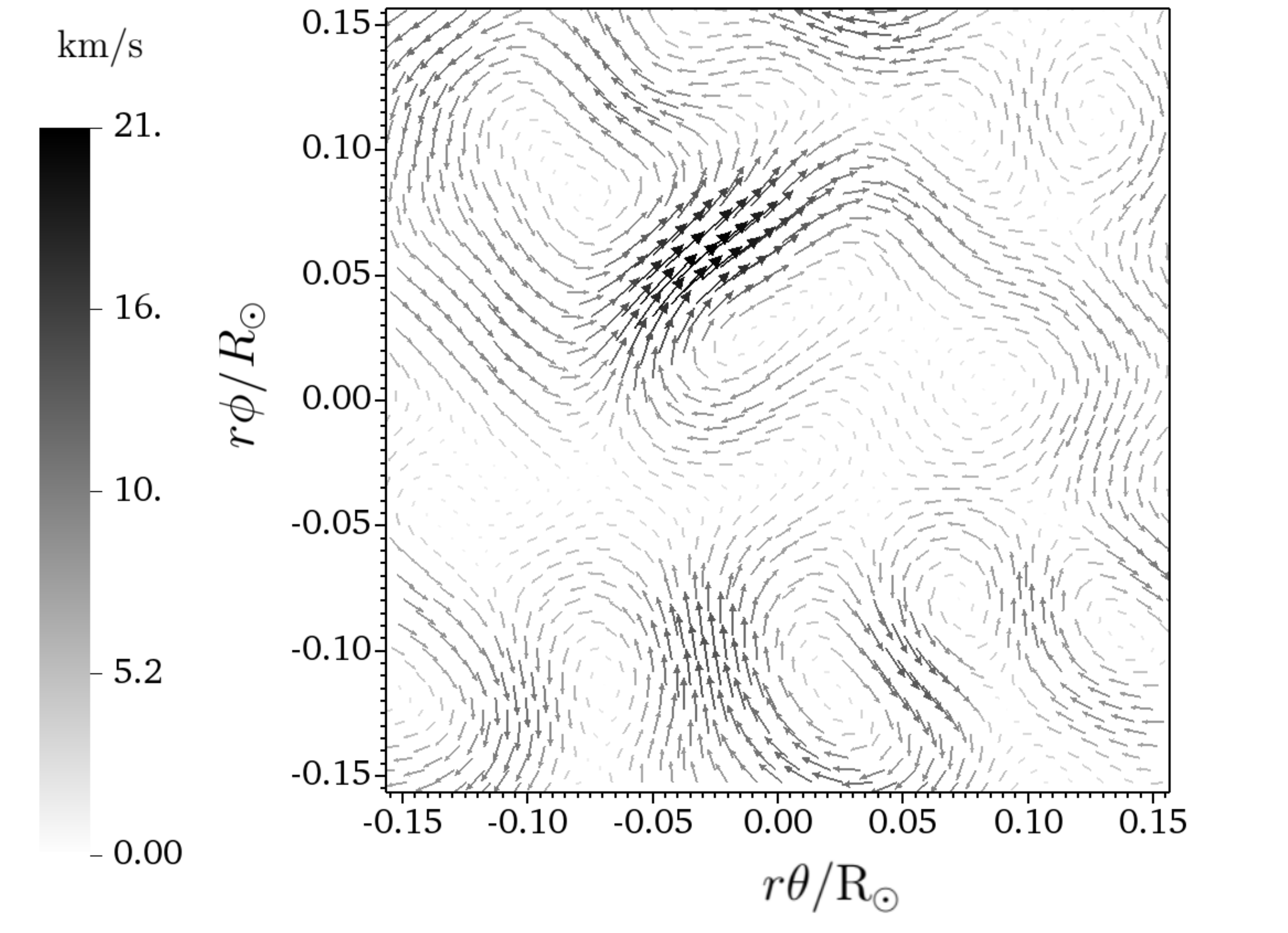}  
        \caption{Spherical slice showing a single realization of the perpendicular velocity vector field at the bottom radial boundary. Note the solenoidicity of the driving velocity.}
        \label{driver}
\end{figure}

The wave driver acts continuously at the bottom boundary, mimicking the omnipresent propagating transverse waves observed to exist in coronal open-field regions \citep{2007Sci...317.1192T,2015NatCo...6E7813M}. The  wave driver adds a time and space-varying solenoidal and purely angular velocity field at selected modes, following an Ornstein-Uhlenbeck (OU) process, which is a well-defined stochastic process with a finite autocorrelation time, adapted from \citet{2010A&A...512A..81F}. We set the autocorrelation time $(\tau_d \approx 350\ \mathrm{s})$, as well as the rms velocity $(v_\mathrm{RMS} \approx 15\ \mathrm{km\ s^{-1}})$ close to their mean values for the imaged waves in polar coronal plumes \citep{2015NatCo...6E7813M}. The driver stirs all transverse waves with wave front wavenumbers up to $8\pi/L_{box}$, where $L_{box} = 220\ \mathrm{Mm}$ is the arc-length of the numerical domain at the bottom radial boundary. This corresponds to characteristic perpendicular scales of $\approx 9-35\ \mathrm{Mm}$, where the upper limit is around the supergranular scale \citep{2018LRSP...15....6R}. All modes have equal energy (flat spectrum). This is justified by the presence of a `bump' in the energy spectrum of the observed transverse waves at the position of the mean wave period \citep{2015NatCo...6E7813M}. The wavenumber of the stirred modes represents the transverse correlation length or perpendicular scales of the wave energy distribution. The choice for the maximal wavenumber is somewhat arbitrary, as the transverse correlation length of Alfv\'enic waves is not known in the lower corona. For a realization of the wave driver at the bottom boundary, see Fig.~\ref{driver}.
\par 
Besides the boundary condition for the velocity field, represented by the driver, at the bottom radial boundary the density and pressure are fixed, while the radial momentum is continuous. The continuity of the radial momentum is regulating the mass loss through the solar wind. The magnetic field is extrapolated in the boundary in a divergence-free manner. At the top radial boundary, all conservative variables are set to continuous, i.e. zero-gradient extrapolation. Numerical tests show no significant reflection on this boundary, aided also by the supersonic and super-Alfv\'enic wind at this radius. As previously hinted, the angular boundaries are periodic. \par  
The numerical domain consists of $1024 \times 256^2$ cells distributed uniformly throughout the quasi-spherical grid. Based on convergence studies, the number of cells along the radial direction are sufficient to not cause significant numerical damping of the outward-propagating waves. 

\section{Results and Discussion}
\label{three}

\begin{figure}[h]
    \centering     
        \includegraphics[width=1.0\textwidth]{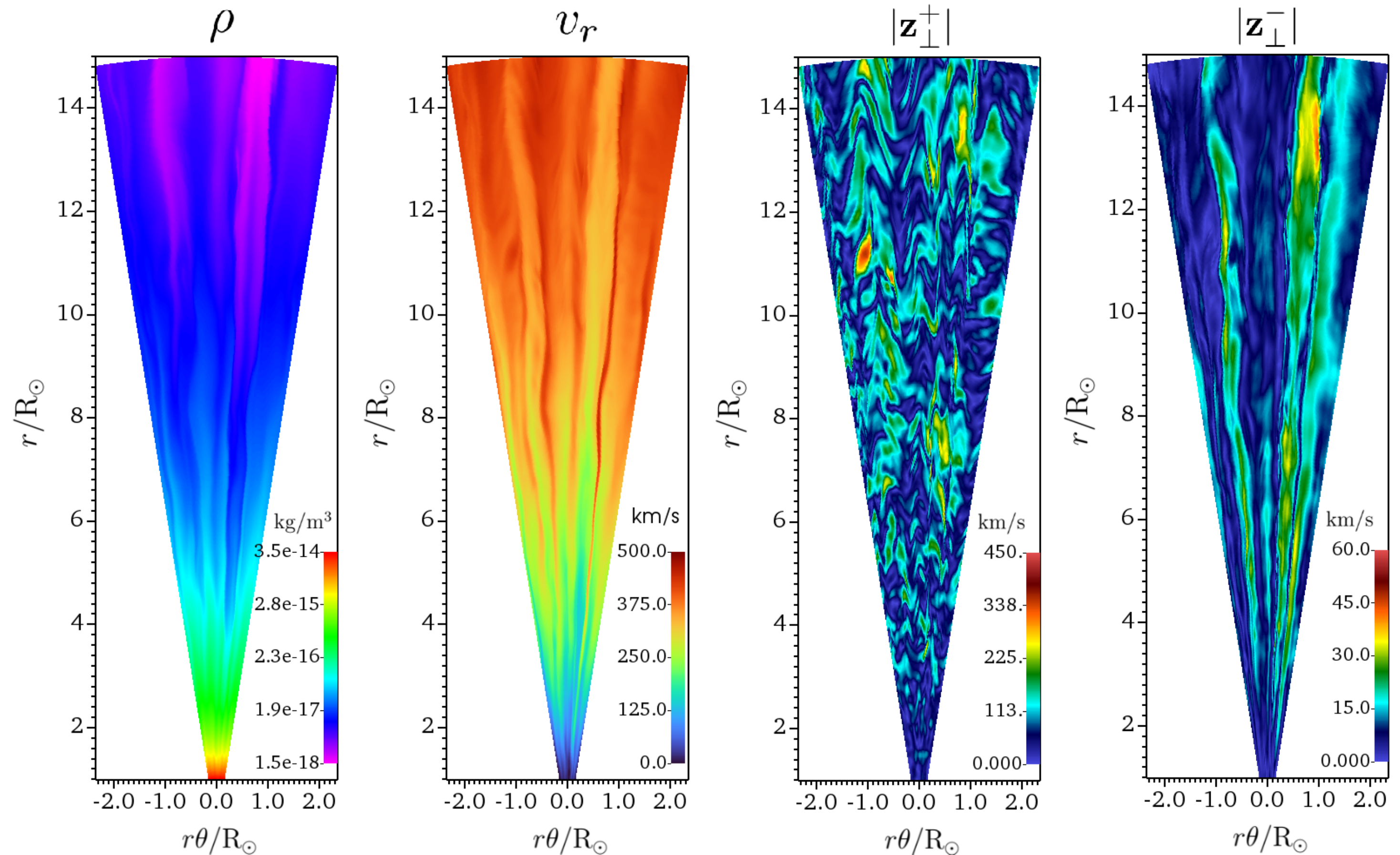}  
        \caption{Snapshots of density ($\rho$), radial velocity ($v_r$), and the perpendicular Els\"{a}sser variables $|\mathbf{z}^+_\perp|$ and $|\mathbf{z}^-_\perp|$, respectively (from left to right), in a slice at $\phi = \pi/2$, $2\ t_A$ after the start of the simulation. In the online version of the paper the figure is animated from $t=0$ to $t=4\ t_A$.}
        \label{sliceZ}
\end{figure}

\begin{figure}[h]
    \centering     
        \includegraphics[width=1.0\textwidth]{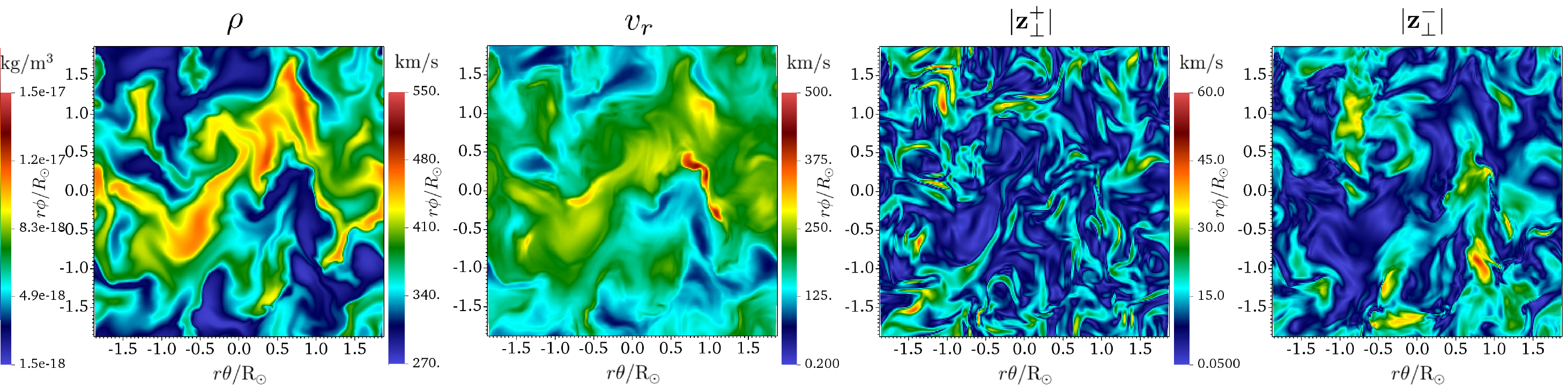}  
        \caption{Same as in Fig.~\ref{sliceZ}, but for a spherical slice at $r = 12\ \mathrm{R}_\odot$. Compare to Fig.~\ref{initcond} to estimate the deformation caused by the driven waves. In the online version of the paper the figure is animated from $t=0$ to $t=4\ t_A$.}
        \label{spslice}
\end{figure}

\begin{figure}[h]
    \centering 
       \begin{tabular}{@{}cc@{}}
        \includegraphics[width=0.5\textwidth]{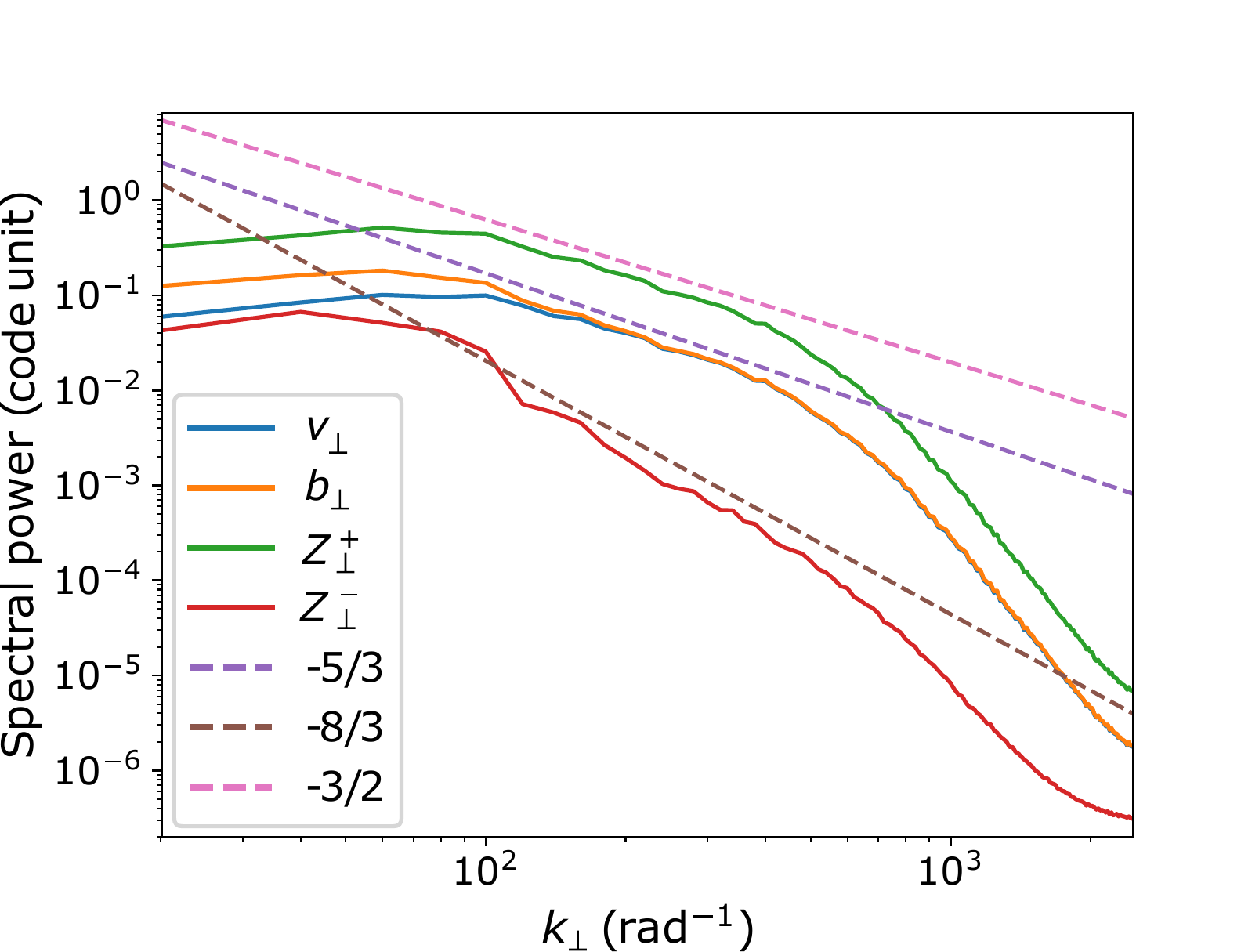}  
        \includegraphics[width=0.5\textwidth]{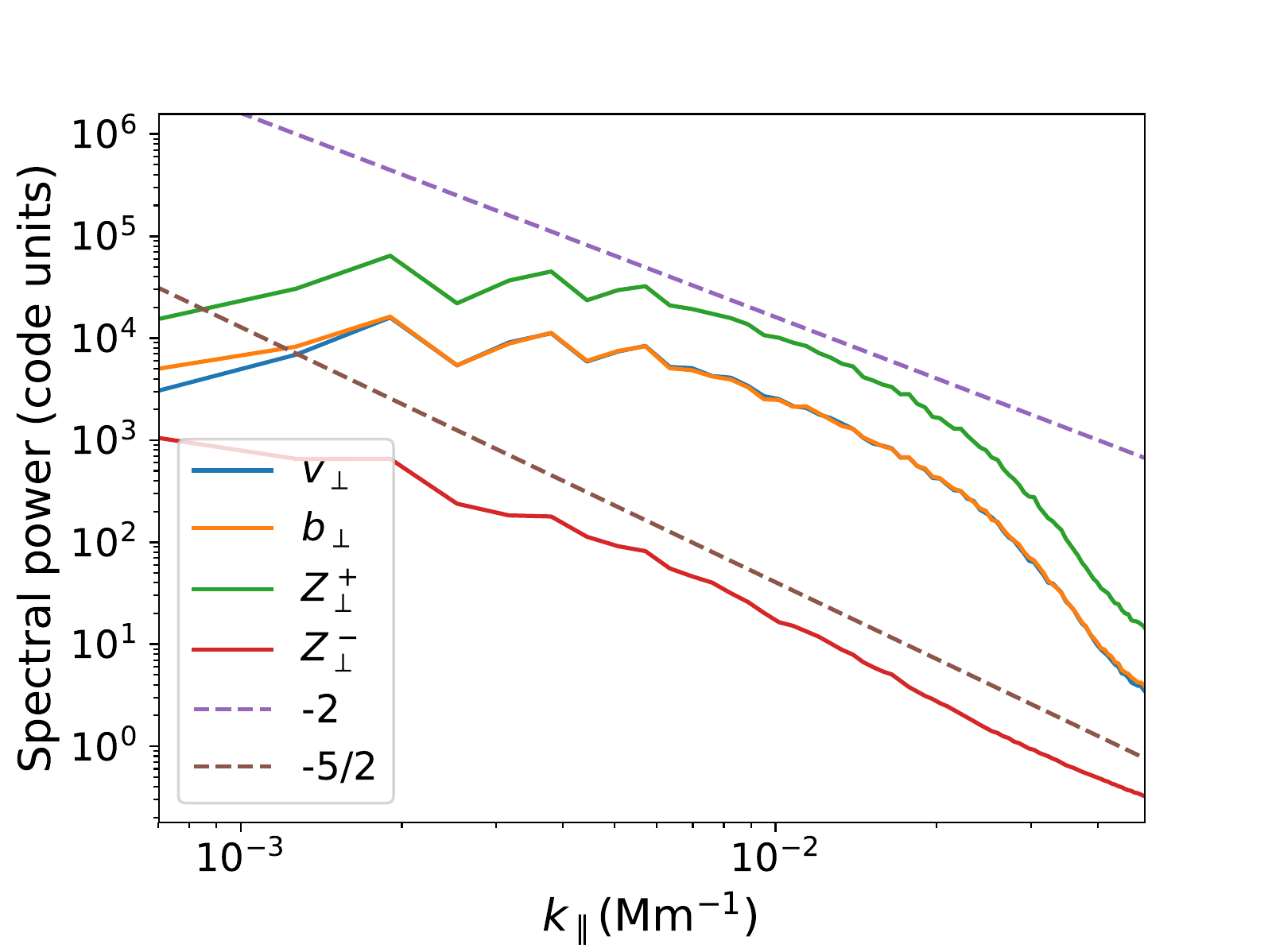} \\
       \end{tabular}    
        \caption{Power spectra of the perpendicular components of the main variables of interest for the inhomogeneous run. \textit{Left}: Perpendicular spectra averaged over multiple spherical slices in the spatial domain from $13\ \mathrm{R}_\odot$ to $15\ \mathrm{R}_\odot$. \textit{Right}: Parallel spectra averaged over multiple radial lines along the whole radial domain. Dashed lines show power laws with the corresponding indices on the legend. Spectra  at $t \approx 4\ t_A$.}
        \label{spectra}
\end{figure}

\begin{figure}[h]
    \centering
       \begin{tabular}{@{}cc@{}}
        \includegraphics[width=0.5\textwidth]{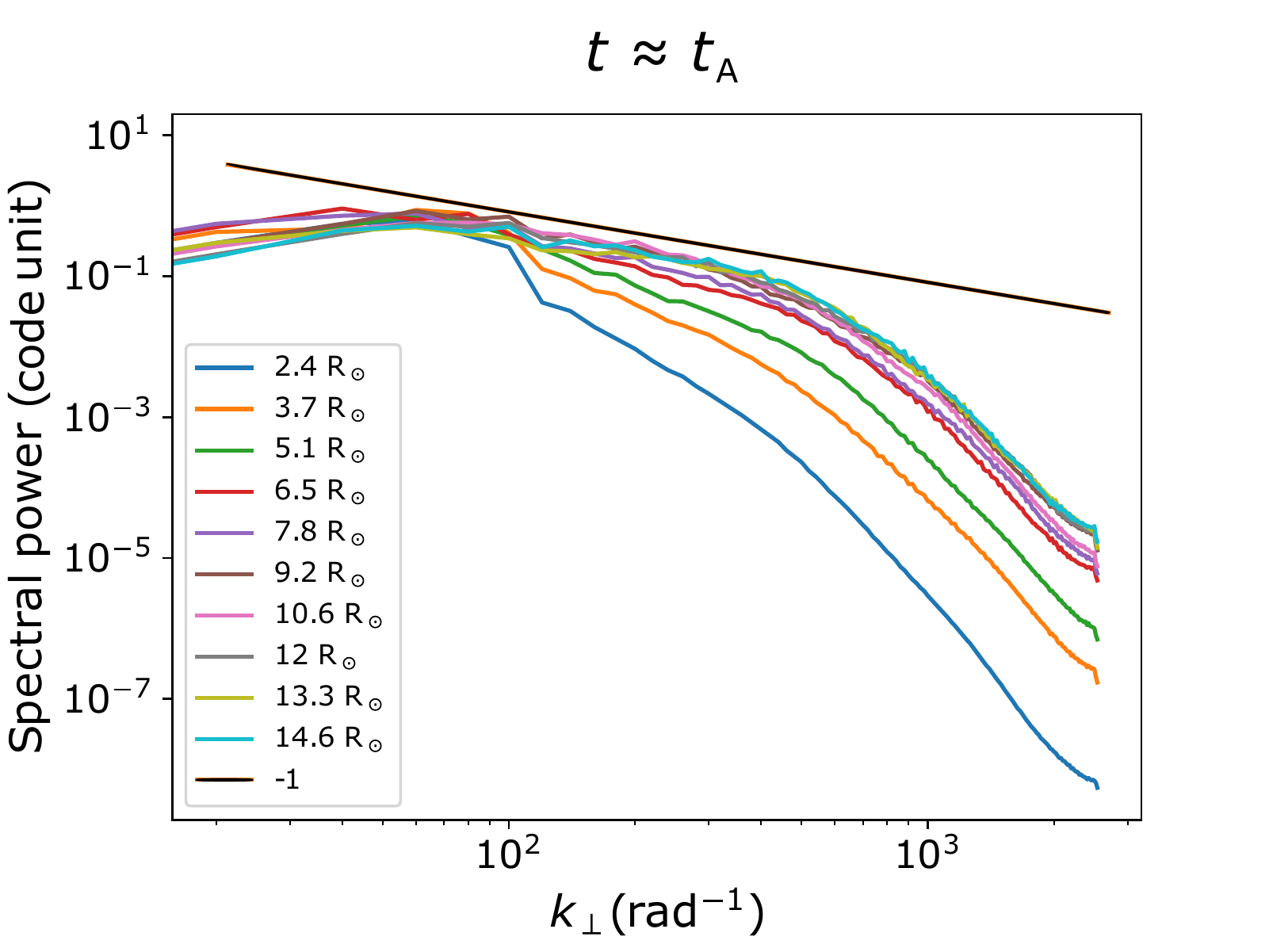}  
        \includegraphics[width=0.5\textwidth]{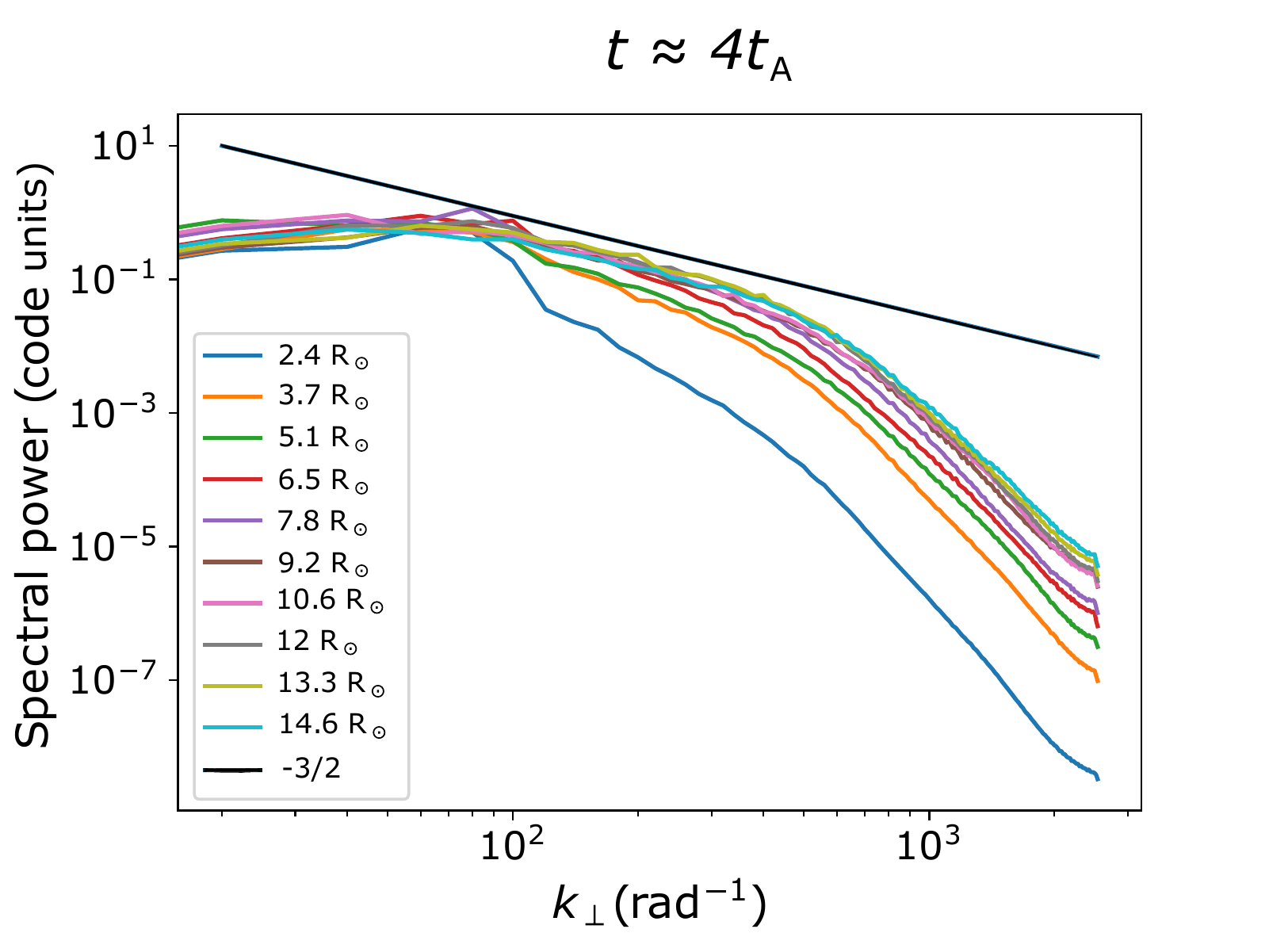} \\
       \end{tabular}  
        \caption{Perpendicular power spectra of $\mathbf{z}^+_\perp$, averaged over multiple slices in the vicinity of the radius indicated on the legend, for the inhomogeneous setup. Also indicated are power laws with corresponding indices on the legend. \textit{Left}: at $t \approx t_A$ \textit{Right}: at $t \approx 4 t_A$}
        \label{spect_evol}
 \end{figure}

\begin{figure}[h]
    \centering     
        \includegraphics[width=0.5\textwidth]{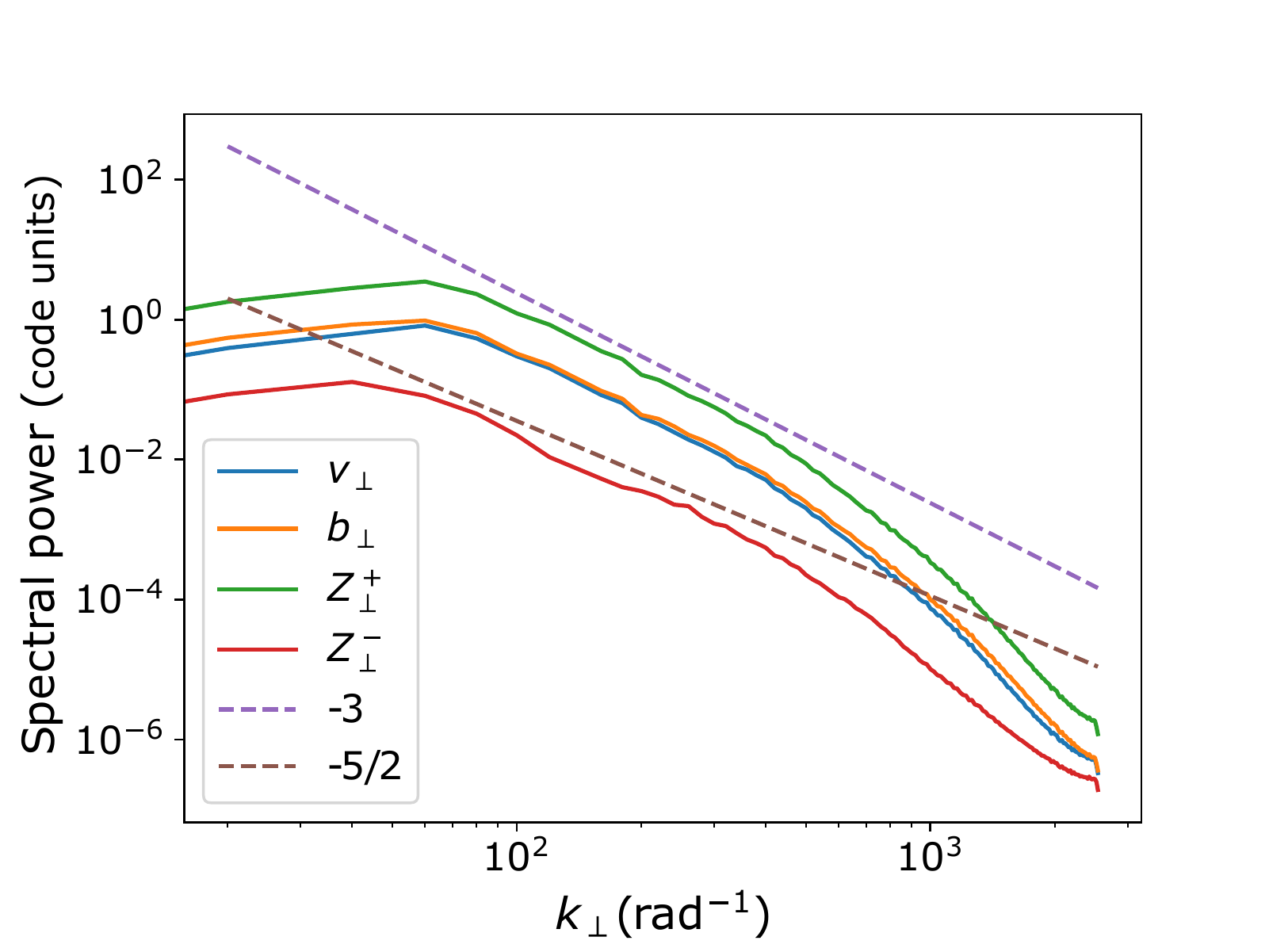}  
        \caption{Power spectra of the perpendicular component of main variables of interest in the homogeneous setup, averaged over multiple slices from $13\ \mathrm{R}_\odot$ to $15\ \mathrm{R}_\odot$, at $t \approx 4t_A$. }
        \label{homogen}
\end{figure}

\begin{figure}[h]
    \centering     
        \includegraphics[width=0.8\textwidth]{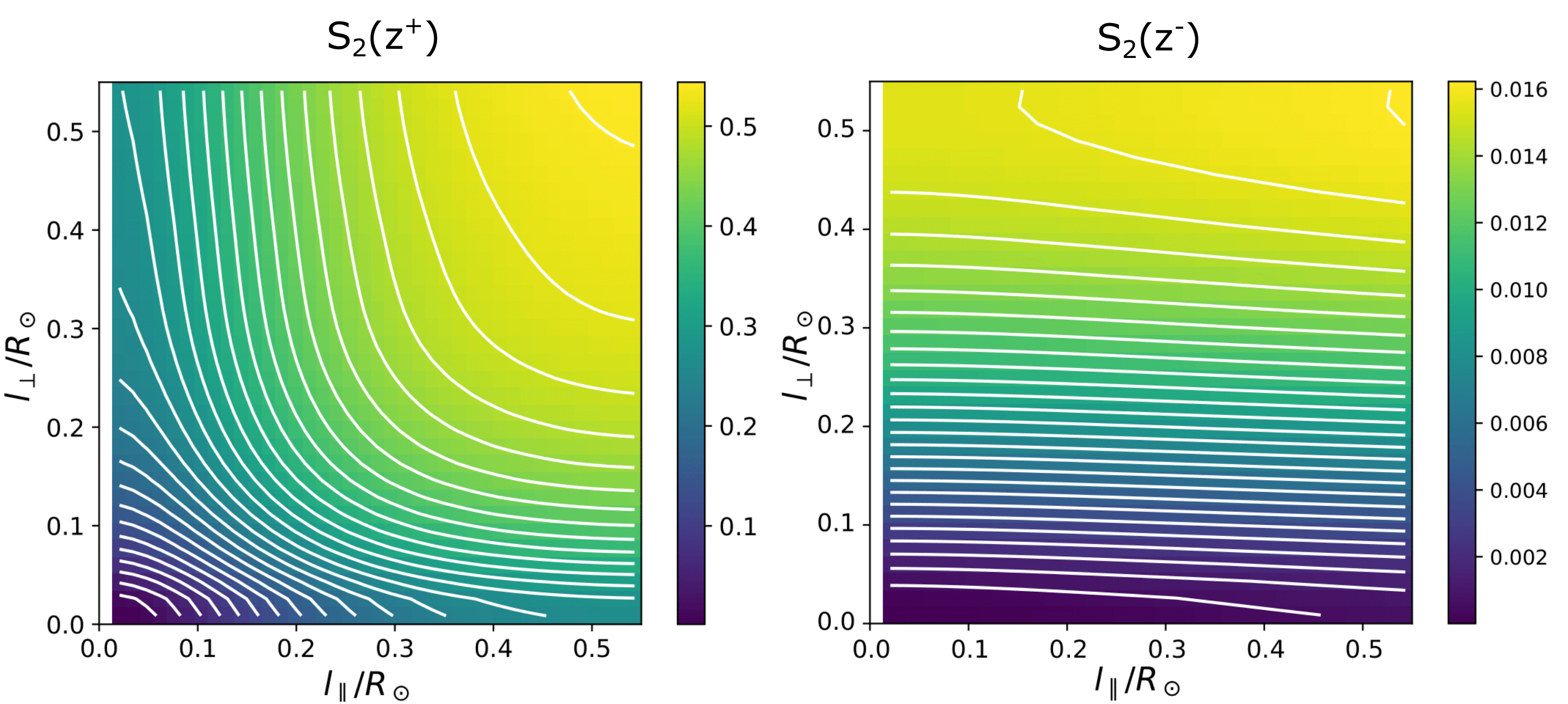}  
        \caption{Two-dimensional second-order structure function ($S_2(z^\pm) = \left\langle(z^\pm(\mathbf{r})-z^\pm(\mathbf{r+l})^2\right\rangle$, averaged over spatial dimensions) of $z^+$ (\textit{Left} panel), and $z^-$ (\textit{Right} panel), around $r = 14\ \mathrm{R}_\odot$, at the end of the simulation. Units are code units.}
        \label{struct}
\end{figure}

As specified in the previous Section, we perform an inhomogeneous simulation, and additionally a homogeneous simulation for comparison between the two cases. The only difference between these simulations is the temperature map at the bottom $T_b$ in Eq.~\ref{Gaussiantemp}. Unless specified otherwise, the following description refers to the inhomogeneous run. The simulation is run for $\approx 10$ hr, while the perpendicularly-averaged Alfv\'en transit time through the simulation domain in the radial direction is $t_A \approx 2.5$ hr. Note that the transit time is different on different field lines. The stochastic driver is continuously acting at the bottom radial boundary. Pure Alfv\'en waves are restricted to Alfv\'en speed isosurfaces in the inhomogeneous setup considered. As these are not specifically driven by the random driver, most of the driven wave power is residing in  propagating transverse displacements of the inhomogeneous structures, that is, kink waves. In Fig.~\ref{sliceZ} and Fig.~\ref{spslice}, snapshots of selected variables in radial and spherical slices are presented, respectively. 
Some immediate observations can be drawn from these snapshots. The evolution appears to be smooth radially when viewed through the density and radial velocity, while perpendicularly it is more fragmented. The velocity amplitude of the driven waves is increasing with radial distance, which is approximately in agreement with the WKB-like $v_\perp \propto \langle\rho^{-1/4}\rangle$ dependence \citep{2001A&A...374L...9M}. Note that the WKB approximation does not take into account many effects that change the amplitude of the outward-propagating waves, such as reflection, wave damping, mode conversion, nonlinear effects, and so on. Instead, it represents an upper bound of the amplitude. Nevertheless, while at the bottom radial boundary rms wave amplitudes are $\approx 0.015\ \mathrm{M_A}$, where $\mathrm{M_A}$ is the Alfv\'en Mach number, by $15\ \mathrm{R_\odot}$ the wave amplitudes are up to $\approx 0.5\ \mathrm{M_A}$ and supersonic. Therefore, the driven waves are expected to undergo a strongly nonlinear evolution. Moreover, the nonlinearity of kink waves is not set directly by $\mathrm{M_A}$. Instead, kink waves are fully nonlinear already for displacements of the same order as the radius of the supporting inhomogeneous structure, e.g. flux tube \citep{2014SoPh..289.1999R}. In this sense, in slender flux tubes, kink waves behave nonlinearly already for velocity amplitudes of $\approx 0.02\ \mathrm{M_A}$. The Els\"{a}sser variable $\mathbf{z}^+_\perp \equiv \mathbf{v_\perp} + \frac{\mathbf{b}_\perp}{\sqrt{\mu \rho}} $ appears to be dominant. Note that  $\mathbf{z}^+_\perp$ would correspond to the anti-sunward propagating Alfv\'en wave in a perpendicularly homogeneous plasma. However, as explained in the Introduction, waves in an inhomogeneous plasma such as kink waves, manifest as co-propagating Els\"{a}sser fields. That is, the Els\"{a}sser variables no longer separate exactly between sunward or anti-sunward propagating waves. Neverteless, as kink waves are still highly Alfv\'enic transverse waves, their anti-sunward propagation direction clearly renders $\mathbf{z}^+_\perp$ their dominant Els\"{a}sser component. The $\mathbf{z}^+_\perp$ component appears patchy radially, in accordance with the finite autocorrelation time of the velocity driver at the bottom. Note as well the oblique wave fronts, especially noticeable in Fig.~\ref{sliceZ} at $r \approx 14\ \mathrm{R_\odot}$ and $r\theta \approx -1\ \mathrm{R_\odot}$. Angular gradients are noticeable across the same location, in both in density and radial velocity. While this phenomena is reminiscent of linear phase mixing \citep{1983A&A...117..220H}, in this case the wave perturbation direction can also be along the Alfv\'en speed gradient, which leads to similar wave front bending \citep{1998JGR...10323691G}, while the nonlinear evolution leads to the self-cascade of waves, or uniturbulence \citep{2019ApJ...882...50M}. 
A striking feature in Fig.~\ref{sliceZ} is the appearance of $\mathbf{z}^-_\perp \equiv \mathbf{v_\perp} - \frac{\mathbf{b}_\perp}{\sqrt{\mu \rho}}$, displaying radially elongated structures, in contrast with $\mathbf{z}^+_\perp$. Note that $\mathbf{z}^-_\perp$ has multiple sources in the current setup. It can represent the minority component of the anti-sunward proapgating kink waves, reflected Alfv\'en and kink waves, evolving non-propagating structures, a build-up of residual energy, directly injected by the driver, and so on. Numerical tests show that the stochastic driver is mainly responsible for the direct injection of the strong $k_\parallel = 0$ component. Additionally, it can represent the generation of a $k_\parallel = 0$ condensate in residual energy \citep{2009PhRvL.103v5001B}, which is consistent with the net magnetic energy build-up in the simulation \citep{2011ApJ...740L..36W}. Nonlinear interactions can lead to a build-up of magnetic energy \citep{2013PhPl...20g2302H}. The $k_\parallel = 0$ component of $\mathbf{z}^-_\perp$ has a similar appearance in both the inhomogeneous and homogeneous simulations. There is a pronounced anti-correlation between the background density and the power in $\mathbf{z}^-_\perp$ fluctuations, also noticeable by comparing the different snapshots in Fig.~\ref{sliceZ}. Test runs show that the exact form of the wave driver (e.g. amplitude, number of wave front modes stirred, wave period) influences the appearance of the $k_\parallel = 0$ power. In the inhomogeneous simulation the space-averaged Els\"{a}sser ratio is $r_E \equiv \left\langle(\mathbf{z}^-_\perp)^2\right\rangle/\left\langle(\mathbf{z}^+_\perp)^2\right\rangle \approx 0.047$ at $t = 4\ t_A$. For comparison, in the homogeneous run with the same wave driver, $r_E \approx 0.016$. Furthermore, we carried out a homogeneous test run with a sinusoidal Alfv\'en wave driver, of the same form as in the Appendix (i.e. $k_\perp = 0, z^-=0$). In this test run, $\mathbf{z}_\perp^-$ is created only by reflection on radial gradients, and as $k_\perp = 0$, nonlinear interactions cannot take place between the driven outgoing and reflected Alfv\'en waves  \citep[see][]{2013PhPl...20g2302H}. This test run allows us to quantify the power in $\mathbf{z}_\perp^-$ generated by reflection alone, and in this case $r_E \approx 7.3 \times 10^{-6}$. Therefore, reflection contributes very weakly to the observed power in $\mathbf{z}_\perp^-$. Note, however that in both homogeneous and inhomogeneous simulations with nonlinear interactions present, $r_E$ is increasing in time, and while it possibly saturates, this does not appear to happen within the simulated time. The increased Els\"{a}sser ratio in the inhomogeneous setup compared to the homogeneous setup is possibly caused by the presence of  kink waves, for reasons discussed above. In solar wind measurements with PSP, $r_E$ at 0.17 AU appears to be consistent with a purely reflection-generated model \citep{2020ApJS..246...53C}, in contrast with our findings that $\mathbf{z}^-_\perp$ is only weakly generated by reflection. Note that the linear non-WKB reflection rate is increasing with wavelength, and the estimate of $r_E$ in the observational study is based on a long-wavelength model \citep{2009ApJ...707.1659C}. Furthermore, density perturbations generated by the parametric decay instability \citep[PDI,][]{1963SPhD....7..988G,1978ApJ...219..700G} can greatly enhance wave reflection \citep{2018ApJ...859L..17S,2019ApJ...880L...2S}. In our simulations, PDI does not play a significant role, mostly because of the lower rms velocity of the driver than in previous simulations. We have investigated the strength of the ensuing turbulence. Turbulence is said to be `strong' if the ratio $\zeta$ of nonlinear (e.g., $\mathbf{z}^- \cdot \nabla \mathbf{z}^+$) to linear ($\mathbf{V}_A \cdot \nabla \mathbf{z}^+$) advection is of the order unity, and `weak' if the ratio is much smaller than unity \citep[e.g.][]{1995ApJ...438..763G}. We find that while linear advection dominates when averaged spatially ($\zeta \approx 0.06$), there are numerous fragmented regions with $\zeta \geq 1$, concentrated in intense hotspots, occupying $\approx 4\%$ of the numerical box volume. In Fig.~\ref{spslice}, deformations and small-scale `ripple' generation are clearly visible both in density and radial velocity. Note the more fragmented appearance of $\mathbf{z}^+_\perp$, compared to $\mathbf{z}^-_\perp$. The same is observed in the homogeneous setup. This is not in agreement with previous numerical results, in which the minority Els\"{a}sser component, corresponding to $\mathbf{z}^-_\perp$, appears to be the more fragmented variable \citep[e.g.,][]{2017NatSR...14820M,2019ApJ...880L...2S}. It is possible that this discrepancy can be attributed to the power $k_\parallel = 0$ in $\mathbf{z}^-_\perp$ induced by the stochastic driver, while in the previous simulations driven with sinusoidal waves this is not observed. In Fig.~\ref{spectra} the perpendicular and parallel spectra of the perpendicular components of various variables are shown, as they appear at the end of the simulation time.
Note that here `perpendicular' and `parallel' are with respect to the initial radial magnetic field, and do not account for the local perturbations to the field. Thus, care should be taken when directly interpreting these spectra as a measure of anisotropy, as in the strict sense this must be measured along the perturbed magnetic field \citep[see, e.g.,][]{2011MNRAS.415.3219C}. Nevertheless, in our simulations the radial direction remains a good approximation of the local mean magnetic field direction. First we investigate the perpendicular spectra. As stated previously, $\mathbf{z}^-$ shows much less small-scale structure, and presents an inertial range slope of around $-8/3$. The slope of $\mathbf{z}^+_\perp$ is well approximated by $-3/2$, but also consistent with $-5/3$, due to the limited resolution. Note the strong build-up of negative residual energy ($E_R = v_\perp^2 - b_\perp^2$, where $b$ is in units of speed, i.e. $b = B/\sqrt{\mu \rho}$) at low wavenumbers, while at higher wavenumbers the magnetic and kinetic energy spectra converge. The magnetic field spectrum appears to be steeper than the velocity spectrum. A steeper magnetic power spectrum is also observed in homongeneous numerical simulations \citep[e.g.,][]{2011ApJ...741L..19B}, and in-situ in the solar wind \citep[e.g.,][]{2013ApJ...770..125C}, but not at 0.17 AU, where both the magnetic field and velocity spectra scale as $-3/2$ \citep{2020ApJS..246...53C}. The parallel spectral indices for both velocity and magnetic field perturbations are steeper, around $-2$, which are in agreement with some theoretical models, e.g. critical balance \citep[e.g.,][]{1995ApJ...438..763G,2012MNRAS.422.3495B}, numerical results \citep[e.g.,][]{2015ApJ...801L...9B,2019PNAS..116.1185M}, and observational evidence in the solar wind \citep[e.g.][]{2008PhRvL.101q5005H,2010MNRAS.407L..31W,2016JPlPh..82f5302C}. On the other hand, NI MHD theory predicts a $-5/3$ parallel scaling \citep{2017ApJ...835..147Z,2020ApJ...900..115Z}, which also appears to be consistent with solar wind observations, including by WIND \citep{2019ApJ...887..160T} and PSP \citep{2020ApJ...898..113Z}.
In Fig.~\ref{spect_evol}, the evolution of the $\mathbf{z}^+_\perp$ spectra along the radial coordinate and in time is shown;
The left panel shows the spectra around the time the first wave front leaves the simulation domain, while the right panel shows the spectra at the end of the simulation time. Note the gradual increase in the power at higher wavenumbers and flattening of the spectrum of $\mathbf{z}^+_\perp$ with increasing radial distance, towards the stable slope value of $-1$, attained around $8\ \mathrm{R_\odot}$. In contrast, at the end of the simulation time the spectra appear to be more similar at different radial distances, and converging towards a slope of $-3/2$ or $-5/3$. The spectra at $ \approx 2.4\ \mathrm{R}_\odot$ appear to be much steeper than at greater radial distances, and it doesn't show evolution in time. This might be related to the peak value of the Alfv\'en speed, where the radial gradient of the Alfv\'en speed becomes zero, occurring at the same radial distance. For an insight on the origin of the difference between the two panels, we look at the different dynamical timescales that are operating in the simulation. For comparing the different timescales, we assume that all waves have the same frequency, i.e. parallel length scale, valid for both $\mathbf{z}^\pm$. Note that these length scales appear to be different in the solar wind for $\mathbf{z}^\pm$ \citep{2015ApJ...805...63A}, which can have an impact on the nonlinear times \citep[see,][]{2012ApJ...745...35Z}. Linear phase mixing, in the sense described earlier, operates on a timescale $\tau_\mathrm{PM} \sim 1/k_\perp^\mathrm{eq} V_{A\parallel}$ \citep{2018ApJ...859L..17S}, where $k_\perp^\mathrm{eq}$ is the specific perpendicular scale of the background inhomogeneity. The nonlinear time of the self-deformation of kink waves, or uniturbulence, was calculated recently for kink waves propagating on a cylindrical flux tube with discontinous boundary, $\tau_\mathrm{UT} \sim R/V (\zeta + 1)/|\zeta - 1|$ \citep{2020arXiv200715411V}, where $R$ is the tube radius, $V$  is the kink wave velocity amplitude, and $\zeta$ is the density ratio. By approximating $R \approx (k_\perp^\mathrm{eq})^{-1}$ and $V \approx z_k^+/2$ for the highly Alfv\'enic kink wave, this formula can constitute an estimate of the nonlinear self-cascade time, $\tau_\mathrm{UT} \sim (1/k_\perp^\mathrm{eq} \delta z^+)(\zeta + 1)/|\zeta - 1|)$ . Note that this time is the `in-flight' nonlinear time, i.e. in a Lagrangian frame moving with the self-deforming perturbation. In the radially finite and open simulation, however, the focus is on the Eulerian nonlinear time, i.e. at specific radial distances from the solar surface. To understand the difference between the two, consider the following thought experiment: a single, nonlinear kink wave-packet propagating upwards through the corona and solar wind is gradually deforming the inhomogeneous structure supporting it (e.g. a higher density cylindrical flux tube). However, subsequently launched wave-packets are propagating through an increasingly deformed inhomogeneous flux tube, which is modifying the eigenfunctions of the propagating waves, which in turn by self-deforming lead to further deformation of the flux tube, and so on. This image is further complicated by the advection of the deformed flux tube by the solar wind. Therefore it is difficult to come up with an Eulerian nonlinear time. On the other hand, the classical weak Alfv\'en wave turbulence of counterpropagating waves operates on a timescale modified by the Alfv\'en effect, which increases the nonlinear time to $\tau_\mathrm{AWT}^\pm \sim k_\parallel v_A / k_\perp^2 {\delta z_k^\mp}^2$, where $k_\parallel$ is the parallel wavenumber, $k_\perp$ is the inverse fluctuation lengthscale, and $\delta z_k^\pm$ is the amplitude of the fluctuations at scale $k_\perp^{-1}$. Note that this nonlinear time is shortened by the faster timescales of linear phase mixing and uniturbulence, leading to effectively higher $k_\perp$. However, as seen in the simulations, most of the power in $\mathbf{z}^-$ is in the form of elongated, $k_\parallel \approx 0$ structures. Therefore, once these structures form, which appears to be immediately following the first wave front, the nonlinear time associated with them should not be $\tau_\mathrm{AWT}$ but shorter as nonlinear interactions become more coherent, $\tau_\mathrm{AN}^\pm \sim 1/k_\perp \delta z_k^\pm$. It is easy to see that the strong interplay between the different mechanisms and their associated timescales renders their separation and/or finding a common nonlinear time nearly impossible. Nevertheless, we can compare the resulting spectra from the homogeneous and inhomogeneous setups, which both display the fast formation of $k_\parallel \approx 0$ power in $\mathbf{z}^-$, as they are driven by the same stochastic driver. In Fig.~\ref{homogen} the perpendicular spectra from the homogeneous setup is shown. 
It is clear that in the homogeneous setup, the total simulation time of $t \approx 4 t_A$ is not sufficient to establish a fully developed turbulent cascade. Therefore, the large-scale perpendicular inhomogeneities appear to speed up the development of a turbulent cascade. In the inhomogeneous simulation, the fastest timescale appears to be that of linear phase mixing, and is thus responsible for the initial power spectra (Left panel of Fig.~\ref{spect_evol}). This shows that for weakly cascading waves linear phase mixing alone can result in a spectral power slope of $-1$. As the simulation progresses, the nonlinear cascade influences the power spectra, resulting in a slope of $-3/2$ or $-5/3$ (Right panel of Fig.~\ref{spect_evol}). Note also that compared to the initial power spectra, there is a weaker variation of the slope with radial distance. There is also no difference in spectra between sub- or super- Alfv\'enic solar wind regions. Note, however that the critical Alfv\'en surface is highly distorted because of the inhomogeneity, with the lowest dips situated at $\approx 11\ \mathrm{R}_\odot$, and the majority of the cross-sectional area super-Alfv\'enic by $\approx 15\ \mathrm{R}_\odot$. Interestingly, the $\mathbf{z}^-_\perp$ spectrum doesn't show the same variation with radial distance and temporal evolution, having always a spectral slope of $\approx -5/2$, dominated by the $k_\parallel = 0$ build-up. This is reflected in the different second-order structure function shapes, presented in Fig.~\ref{struct};
These plots demonstrate that there is more variation in the perpendicular direction than along the radial direction. In calculating the structure function, we assumed that the turbulence is isotropic in the perpendicular directions. \par
\begin{figure}[h]
    \centering     
        \includegraphics[width=0.8\textwidth]{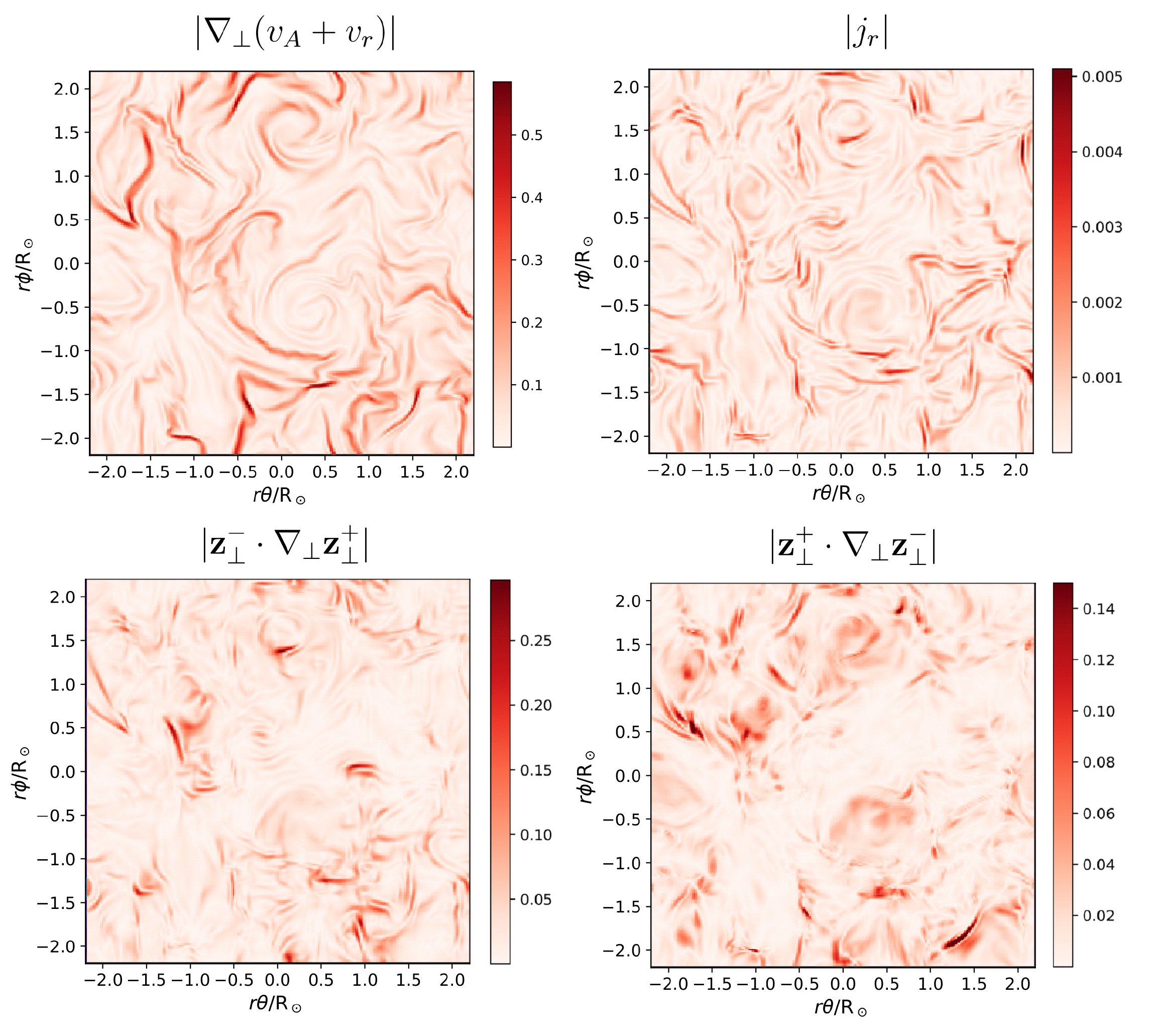}  
        \caption{Snapshots of Doppler-shifted Alfv\'en speed perpendicular gradient (\textit{Top Left}), radial electrical current (\textit{Top Right}), and the perpendicular nonlinear advection terms for $z^\pm$ (\textit{Bottom}), in spherical slices at $r = 12\ \mathrm{R}_\odot$, at $t\approx 4 t_A$. Units are code units.}
        \label{nonlin_grad}
\end{figure}

 
In Fig.~\ref{nonlin_grad}, variables relevant to the nonlinear cascade are shown in a spherical cross-section. 
The bursty appearance of nonlinear advection, concentrated in intense hotspots, is in agreement with the findings above that strong nonlinearity is occupying a small percentage of the total volume.
The most important observation is the correlation between perpendicular Alfv\'en speed gradients ($\nabla_\perp v_A$), the radial current $j_r$, and the nonlinear advection terms ($z^\pm_{NL}$, defined in Fig.~\ref{nonlin_grad}). We use the `correlate' function of the \texttt{NumPy}\footnote{https://numpy.org/} Python package between the flattened (decomposed into 1D arrays) 3D subsets of the data from $r=12\ \mathrm{R_\odot}$ to $r=15\ \mathrm{R_\odot}$. In this sense, we find the overall cross-correlation between two variables $A$ and $B$ in the whole 3D subset, and we denote this operation as $A \star B$. The calculated cross-correlations are the following: $\nabla v_A \star |j_r| \approx 0.23$; $\nabla v_A \star z^+_{NL} \approx 0.17$; $\nabla v_A \star z^-_{NL} \approx 0.23$; $|j_r| \star z^+_{NL} \approx 0.47$; $|j_r| \star z^-_{NL} \approx 0.15$; for comparison, the Alfv\'en correlation is, e.g., $v_\phi \star b_\phi \approx 0.9$. Surprisingly, in the homogeneous setup, similar values are found for the cross-correlations, albeit from much lower amplitude relative perpendicular Alfv\'en speed perturbations. Note that perpendicular Alfv\'en speed inhomogeneities are induced nonlinearly in the homogeneous setup. These might originate from ponderomotive forces, and from the various nonlinear interactions of the driven waves \citep{2018ApJ...859L..17S}. Therefore, it is not clear whether these correlations are specific to inhomogeneous conditions or to compressible homogeneous turbulence in general. There is an especially strong correlation between the nonlinear advection of $\mathrm{z}^+$ and $j_r$. Taking into account the correlation with Alfv\'en speed gradients as well, it is likely that resonant absorption, i.e. linear resonant coupling of kink and Alfv\'en waves, is taking place in these regions \citep[e.g.,][]{1974PhFl...17.1399C,1978ApJ...226..650I,2008ApJ...687L.115T}. Component magnetic reconnection sites are also possible in the resulting thin current sheets \citep[e.g.,][]{2003SoPh..214..107L}. It is not clear how all these effects are impacting the nonlinear cascade of kink waves. It is important to point out here that the nonlinear advective terms plotted in Fig.~\ref{nonlin_grad} are not the only nonlinear terms that might contribute to the turbulent dynamics, as noted in the Introduction. Inhomogeneities and compressibility allow for multiple additional nonlinear terms \citep[e.g.][]{1987JGR....92.7363M,1989GeoRL..16..755Z,2019ApJ...882...50M}. We have investigated the appearance of one additional nonlinear term, present due to density inhomogeneities \citep[see][and the Appendix]{2019ApJ...882...50M}, which is on the same order of magnitude as the nonlinear advective terms. Nevertheless, we have opted to show only the nonlinear advective terms in Fig.~\ref{nonlin_grad}, as these are the terms usually associated with turbulence generation.\par
Though not the main focus of this study, there is no significant solar wind heating and acceleration added by the onset of turbulence in our model. This is probably the result of the rms amplitude of the wave driver at the bottom radial boundary. While previous models often employ a driver with rms velocities given by the observed non-thermal line broadening, here we opted for rms velocities of swaying coronal plumes, inferred from direct imaging \citep{2015NatCo...6E7813M}. The amplitude of the former \citep[$\approx 40\ \mathrm{km/s}$, e.g.,][]{1998SoPh..181...91D,2009A&A...501L..15B,2012ApJ...751..110B} is almost three times the rms amplitude of the latter \citep[$\approx 15\ \mathrm{km/s}$,][]{2015NatCo...6E7813M}. However, there is no consensus on the physical nature of the non-thermal line broadening other than it represents unresolved motions \citep[e.g.,][]{2020A&A...639A..21P}. Moreover, even though some solar wind simulations are able to maintain a hot corona and an accelerated solar wind, the dissipation coefficients in these models are orders of magnitude higher than in reality. Still, one may argue that once turbulence develops, the start of the dissipation range (in k-space) is not relevant.

\section{Conclusions} 
\label{four}

We have conducted a first, full 3D compressible MHD simulation on the dynamics of kink waves from the base of the perpendicularly inhomogeneous corona up to $15\ \mathrm{R}_\odot$. The large-scale plasma structuring in our model is aimed to correct for the considerable gap between the state-of-the-art homogeneous models and the observed reality of a highly structured outer solar atmosphere. The \textit{Parker Solar Probe} will deliver, in the near future, unprecedented in-situ measurements of the outer solar corona and pristine solar wind. Therefore, it is high time that solar wind models incorporate the important effects of structuring. \par 
Waves launched by a stochastic driver at the bottom radial boundary propagate anti-sunward and suffer both linear and nonlinear changes.
Plasma structuring across the radial magnetic field is responsible for some of these. As pure Alfv\'en waves can only exist on Alfv\'en speed isosurfaces, which are not selectively driven, most of the wave power is in kink waves; That is, transverse waves which owe their existence to the structuring and are predominantly driven by magnetic tension. Kink waves manifest as propagating transverse displacements of the structures they exist on. \par 
Linear phase mixing, both perpendicular to the perturbation direction \citep{1983A&A...117..220H}, and along perturbation components \citep{1991ApJ...376..355P,1998JGR...10323691G}, is the first mechanism responsible for the population of energy at small scales, after the start of the driving. This is also observed in the homogeneous compressible MHD turbulence simulations of \citet{2018ApJ...859L..17S}. In addition, the dependence of the propagation speed on the amplitude may lead to nonlinear phase mixing of any pure Alfv\'en waves propagating at neighbouring magnetic surfaces \citep{2017ApJ...840...64S}. Resonant absorption, the linear resonant coupling of kink modes and Alfv\'en waves and their subsequent phase mixing is also contributing to the redistribution of energy \citep{2015ApJ...803...43S,2020ApJ...893..157E}. The phase-mixed waves show a power law spectrum scaling of $-1$, at scales smaller than the driven scale. A scaling of $-1$ is also observed in the fast solar wind, for frequencies below $\approx 10^{-3}\ \mathrm{Hz}$ at 1 AU, with the breaking value shifting to lower frequencies with radial distance \citep{2013LRSP...10....2B}. This so-called 1/f range is considered the driving scale for the turbulent cascade, although its origin is not well understood and still under debate \citep{2012ApJ...750L..33V,2018JPlPh..84a9006C,2018ApJ...869L..32M}. Our results might preliminarily indicate that the 1/f range could also be the result of phase-mixed, long wavelength waves with a slow nonlinear evolution, their gradual nonlinear cascade with radial distance indicated by the aforementioned shifting of the breaking value. However, more research is needed to determine whether this is the case. \par
Waves propagating on structured plasma, such as kink waves, have unique properties compared to pure Alfv\'en waves which are  important for their nonlinear evolution. Unlike pure Alfv\'en waves, these waves perturb both Els\"{a}sser variables as they propagate. This leads to a coherent self-cascade of the waves. Although this phenomenon was observed in previous, simpler numerical simulations \citep{2017NatSR...14820M,2019ApJ...882...50M}, due to the increased complexity of the present model it is difficult to identify directly the self-cascade. Therefore, in the Appendix, we present the results of simplified simulations, with a single, cylindrical `plume' inhomogeneity. In these simulations the evolution of kink waves can be followed more clearly, providing further evidence to uphold our conclusions as presented here. \par  
We have compared the results of the inhomogeneous simulation to a homogeneous, but otherwise identical simulation with a similar stochastic driver. In both setups, the stochastic driver induces a strong $k_\parallel \approx 0$ power in the minority Els\"{a}sser variable $\mathbf{z}^-$, with a steep (although short, due to resolution limitations) inertial range spectra. The presence of kink waves in the inhomogeneous setup leads to $\approx 3$ times higher Els\"{a}sser ratio than in the homogeneous setup. In a test homogeneous simulation driven with $k_\perp = 0$ `slab' Alfv\'en waves, which do not satisfy the condition for nonlinear interactions, the power in $\mathbf{z}^-$ represents only the reflected Alfv\'en waves along radial Alfv\'en speed gradients, and is orders of magnitude smaller than in the homogeneous setup with stochastic driver. Therefore, there is only a negligible power of $\mathbf{z}^-$ generated by reflection. However, nonlinear interactions lead to a build-up of negative residual energy, and this leads to a steady increase of the Els\"{a}sser ratio over time. The appearance of $\mathbf{z}^-$ differs greatly from that of $\mathbf{z}^+$, and is in agreement with the in-situ measured much longer correlation time for $\mathbf{z}^-$ compared to the one for $\mathbf{z}^+$ \citep[e.g.,][]{2020ApJS..246...53C}. However, the steep spectrum of $\mathbf{z}^-$ is not in agreement with observations. Despite the largely similar appearance of $\mathbf{z}^-$ in the homogeneous and inhomogeneous setups, the energy spectra in the inhomogeneous setup evolves towards the well-known $-3/2$ or $-5/3$ slope after around $2\ t_A$, while in the homogeneous setup we find a slope of $\approx -3$ after $4\ t_A$. This indicates that nonlinear interactions in the homogeneous setup are considerably weaker, and that the simulated time is not enough to set up a well-developed turbulent cascade. We conclude from this observation that nonlinear interactions of the outgoing waves with the $k_\parallel \approx 0$ component of $\mathbf{z}^-$ are not the cause of the developing turbulence in the inhomogeneous setup. This hints that the nonlinear self-cascade of kink waves, or `uniturbulence', might play a leading role in setting up the turbulent cascade. In previous solar wind models, the perpendicular structuring of the solar corona and low solar wind are not self-consistently included as inhomogeneities in the background plasma, as described in the Introduction. Instead, a `transverse correlation length' is usually considered for the fluctuations, which is aimed to account for the structuring. This approach is actually identical to our homogeneous simulation, such that the stochastic driver has a specific correlation length transverse to the magnetic field, but otherwise the plasma is perpendicularly homogeneous. In this paper we show that considering only a transverse correlation length of fluctuations is insufficient in accounting for the effects of transverse background inhomogeneities. These need to be included self-consistently, as they allow for the existence of different types of waves, such as kink waves, which have a strong effect on the resulting turbulent evolution.\par 
Many of the nonlinear effects that have considerable impact on the dynamics of kink waves depend on the amplitude of the driven waves at the bottom radial boundary. In our simulation, the driving amplitude is adjusted to the imaged mean velocity amplitude of swaying plumes, and it is 2 to 3 times smaller than the one inferred from non-thermal line broadening. Generally, nonlinear times are inversely proportional to the velocity amplitude of the waves, therefore simulations with higher driving amplitudes might result in a different picture. The parametric decay instability, which is negligible in our simulation, might become dominant for higher amplitudes, leading to strong, localized density enhancements and enhanced wave reflection \citep{2019ApJ...880L...2S}. \par
Although it appears clear that the self-cascade of kink waves, or uniturbulence, operates on faster timescales than Alfv\'en wave turbulence, it is worth discussing here about the possible broader implications of this. Most importantly, how is the different cascade phenomenology impacting the fully developed turbulence in the statistical sense. Are there specific signatures of uniturbulence that are essentially different from those of Alfv\'en wave turbulence? That is, signatures that could be detected by single-point measurements in the solar wind. The spectral properties of the developed turbulence in the inhomogeneous model, as they appears here, are essentially not different from those in previous numerical results based on Alfv\'en wave turbulence, except for the spectra of the minority Els\"{a}sser variable, influenced by the driver. As mentioned previously, a possible fingerprint of the self-cascade is lacking, in the statistical sense. Thus, finding parameters that are unique to this nonlinear cascading channel would be imperative in order to establish its presence in the solar wind, which we leave as a subject of future studies. The correlation of perpendicular Alfv\'en speed gradients, radial current, and the nonlinear advective terms is noteworthy, however the same is observed in the homogeneous simulation, where perpendicular inhomogeneities arise ponderomotively or from other nonlinear sources. \par 
The impact of uniturbulence on the heating rate and solar wind acceleration is an important follow-up study. The wave energy contained in the observed plume swaying appears to be insufficient to raise the average temperature and the wind outflow speed above their base value in the current setup. Therefore, the flow energy available in the nonthermal line broadening might represent the additional energy source required to accelerate the fast wind to the measured values.

\appendix
\section{Single-plume simulations}

\begin{figure}[h]
    \centering     
        \includegraphics[width=1.0\textwidth]{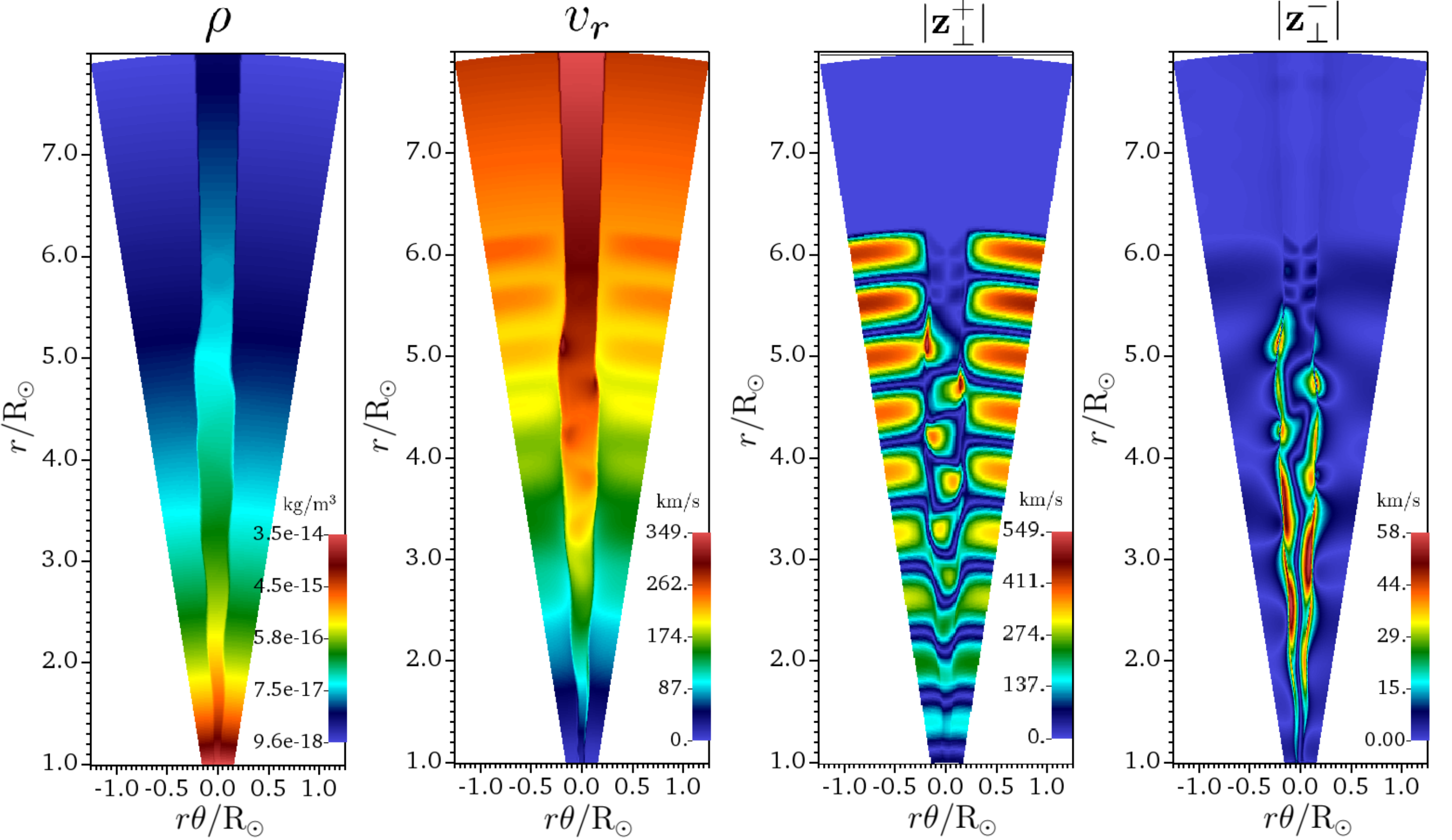}  
        \caption{Snapshots of density ($\rho$), radial velocity ($v_r$), and the perpendicular Els\"{a}sser variables $|\mathbf{z}^+_\perp|$ and $|\mathbf{z}^-_\perp|$, respectively (from left to right), in a slice at $\phi = \pi/2$, $0.75\ t_A$ after the start of the simulation, in the sinusoidally-driven plume simulation.}
        \label{Asinz7}
\end{figure}

\begin{figure}[h]
    \centering     
        \includegraphics[width=1.0\textwidth]{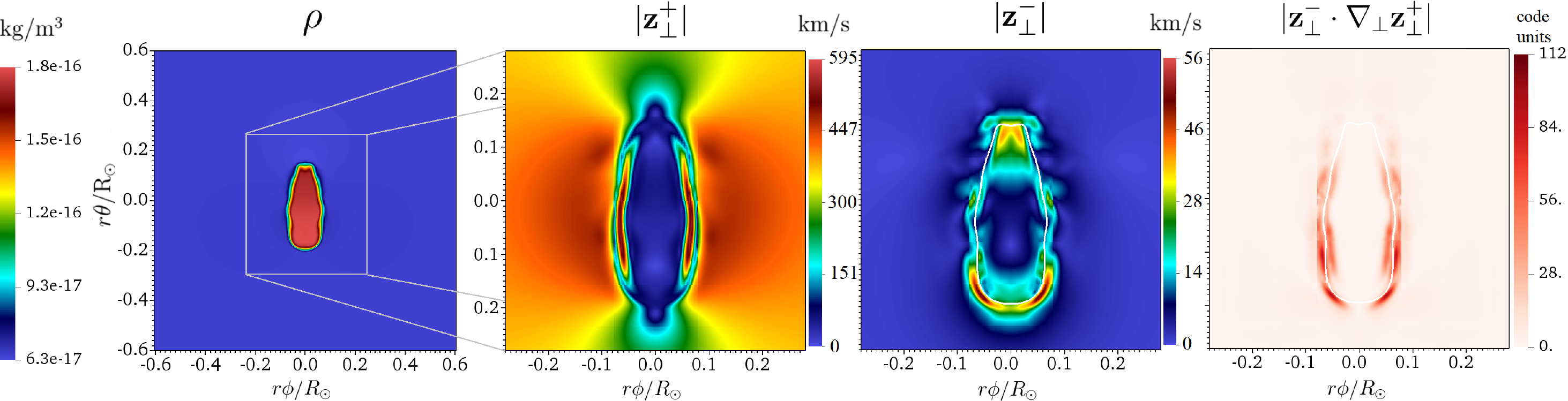}  
        \caption{Snapshots of density ($\rho$), the perpendicular Els\"{a}sser variables $|\mathbf{z}^+_\perp|$ and $|\mathbf{z}^-_\perp|$, and of the nonlinear advection of $\mathbf{z}^+_\perp$, respectively (from left to right), in a spherical slice at $r = 4\ \mathrm{R}_\odot$, $0.75\ t_A$ after the start of the simulation, in the sinunsoidally-driven plume simulation. Note that the three rightmost plots are of zoom-in regions of the first plot, as outlined by the white rectangle.}
        \label{Asinsp7}
\end{figure}

\begin{figure}[h]
    \centering     
        \includegraphics[width=1.0\textwidth]{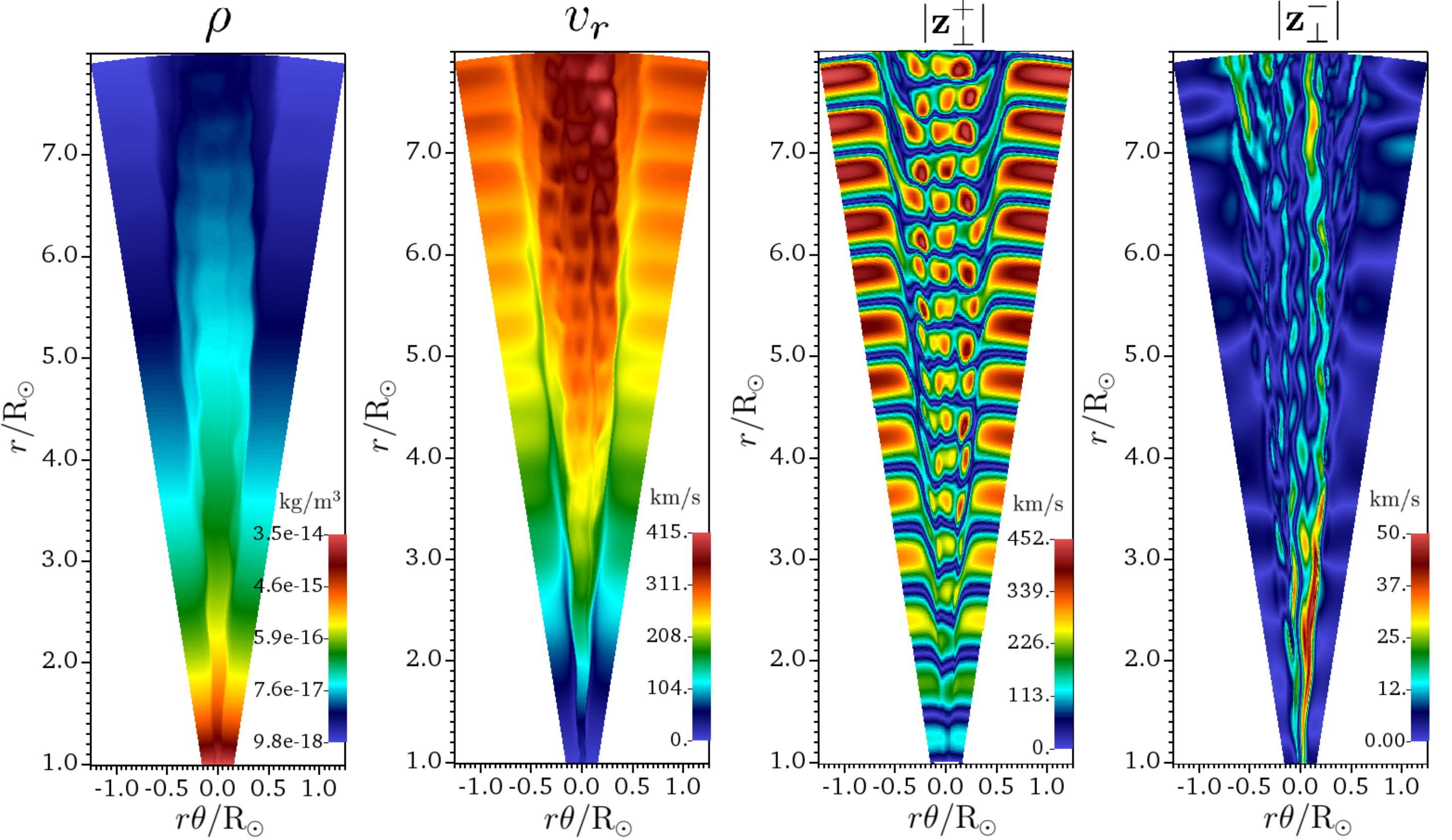}  
        \caption{Snapshots of density ($\rho$), radial velocity ($v_r$), and the perpendicular Els\"{a}sser variables $|\mathbf{z}^+_\perp|$ and $|\mathbf{z}^-_\perp|$, respectively (from left to right), in a slice at $\phi = \pi/2$, $1.5\ t_A$ after the start of the simulation, in the sinusoidally-driven plume simulation.}
        \label{Asinz30}
\end{figure}

\begin{figure}[h]
    \centering     
        \includegraphics[width=1.0\textwidth]{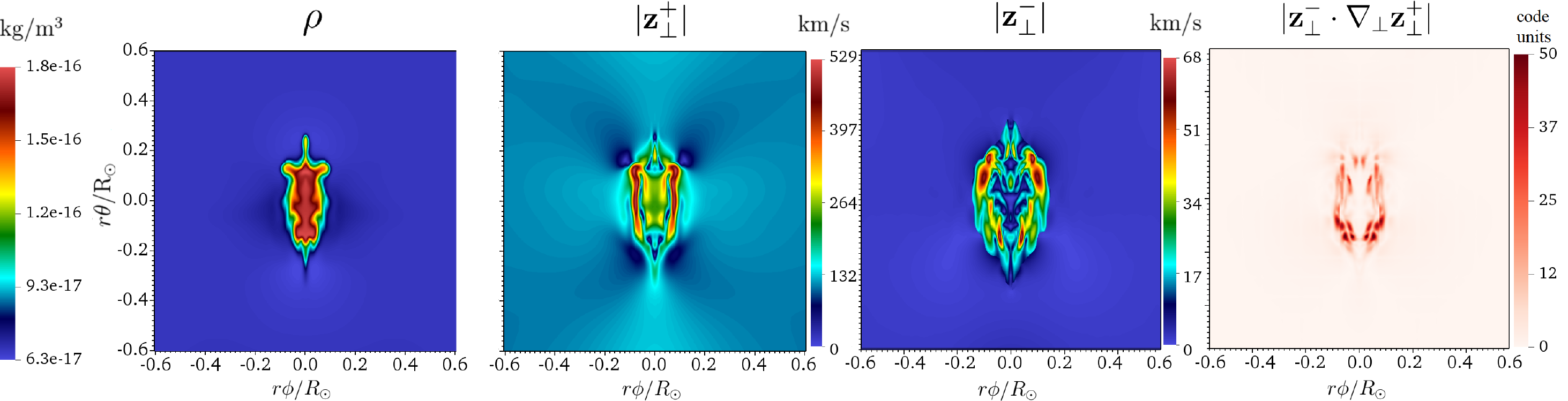}  
        \caption{Snapshots of density ($\rho$), the perpendicular Els\"{a}sser variables $|\mathbf{z}^+_\perp|$ and $|\mathbf{z}^-_\perp|$, and of the nonlinear advection of $\mathbf{z}^+_\perp$, respectively (from left to right), in a spherical slice at $r = 4\ \mathrm{R}_\odot$, $1.5\ t_A$ after the start of the simulation, in the sinunsoidally-driven plume simulation.}
        \label{Asinsp30}
\end{figure}

The inhomogeneous simulation presented in the main part of this paper is populated by randomly positioned and shaped structures, and driven by a stochastic wave driver. This is necessary for a realistic representation and for statistically meaningful analysis. However, the complexity of this simulation hinders the easy identification of the many, both linear and nonlinear processes which contribute to the turbulent picture, and which also act concomitantly. Therefore, in order to clearly reveal these processes and for additional proof of the claims in this paper, such as the dominance of uniturbulence in the inhomogeneous setup, in this Appendix we present the results of simulations containing a single background inhomogeneity. This inhomogeneity consists of a cylindrical higher-density plume, which results from considering a temperature map in Eq.~\ref{GaussianHeat} of the form:
\begin{equation}
T_b(r=1.01\ R_\odot,\theta,\phi) =  
\begin{cases} 
      1.3 T_0, & R\leq  18\ \mathrm{Mm},\\
      T_0, &  R> 18\ \mathrm{Mm},\\
\end{cases}
\label{plume}
\end{equation} 
where $R = 1.01\mathrm{R}_\odot\sqrt{(\theta-\pi/2)^2+(\phi-\pi/2)^2} = 18\ \mathrm{Mm}$ delimits the outer circular edge of the plume centred around $\theta = \phi = \pi/2$. Although Eq.~\ref{plume} is a piecewise function, due to nonzero numerical diffusion the plume boundary is slightly smoothed as the simulation starts. We run two simulations using two different wave drivers at the bottom boundary: A simulation with the stochastic driver described in Section~\ref{two} with identical parameters, and a simulation with a simple, linearly-polarized sinusoidal driver without spatial dependence, i.e. $k_\perp = 0$: 
\begin{equation}
v_\phi(r=1.01\ R_\odot,t) = A\ \mathrm{sin}\left(\frac{2 \pi}{T}t\right),
\label{sindriver}
\end{equation}  
where $A = 30\ \mathrm{km/s}$ is the wave amplitude and $T = 500\ \mathrm{s}$ is the wave period. These single-plume simulations, apart from the aforementioned differences in the temperature map and the sinusoidal driver, are similar to the ones presented in Section~\ref{three}, except that their upper radial boundary is at $r = 8\ \mathrm{R}_\odot$, because of computational resource limitations. The simulations are run for $2 t_A$, where $t_A$ is the Alfv\'en transit time in the radial direction, outside of the plume. In increasing complexity order, we first present the results of the sinusoidal linearly-polarized driver. In Fig.~\ref{Asinz7} and Fig.~\ref{Asinsp7}, we present snapshots of the moment the first driven wavefront reaches up to $\approx 6\ \mathrm{R_\odot}$.
We can already observe several effects. First, note that the driver excites a kinking of the single plume, which is identified as a propagating kink wave, apparent in both the density and radial velocity maps in Fig~\ref{Asinz7}. Analyzing the appearance of $|\mathbf{z}_\perp^+|$ in Fig.~\ref{Asinz7}, it appears that the driver excites a kink wave in the plume, which is surrounded by propagating Alfv\'en modes in the homogeneous region external to the plume. Note that the phase speed of the kink wave is smaller than that of the surrounding Alfv\'en waves \citep[e.g.,][]{1983SoPh...88..179E}, the first observed deformation being at $\approx 5\ \mathrm{R_\odot}$ by the time the Alfv\'en waves reach $\approx 6\ \mathrm{R_\odot}$. The growth of the perturbation amplitude with increasing radial distance is in agreement with the WKB-like $\rho(r)^{-1/4}$ trend \citep{2001A&A...374L...9M}. Linear phase mixing in the perturbation direction (i.e. along $\hat{\theta}$), also results from the difference in phase speeds, as seen in the progressively stronger decorrelation of the kink and Alfv\'en waves with increasing radial distance. While the surrounding $k_\perp = 0$ Alfv\'en waves show minimal $|\mathbf{z}_\perp^-|$ perturbation, mostly due to linear coupling through reflection \citep[e.g.,][]{1980JGR....85.1311H}, the kink waves display a strong co-propagating $|\mathbf{z}_\perp^-|$ component, as discussed previously, mostly situated around the edge of the plume. Other important things to notice are signatures of strong nonlinearity. Note the ponderomotively induced radial velocity perturbations \citep[e.g.,][]{1999JPlPh..62..219V}, both of the kink and the Alfv\'en wave. Although hardly noticeable in Fig.~\ref{Asinz7}, the Alfv\'en waves also steepen nonlinearly. Also note the strong deformation of the initially circular cross-sectional plume, a definite sign of nonlinearity. This is noticeable as the elongation of the plume along $\hat{\theta}$ in both Fig.~\ref{Asinz7} and Fig.~\ref{Asinsp7}. Signs of nonlinearity are also the growing and alternating asymmetry in the perturbation profile of the kink mode along $\hat{\theta}$, apparent in $|\mathbf{z}_\perp^+|$ of Fig.~\ref{Asinz7}. Focusing now on Fig.~\ref{Asinsp7}, resonant absorption, a linear process, is clearly present in $|\mathbf{z}_\perp^+|$, appearing as two thin and strong perturbations across the edges of deformed plume \citep[e.g.,][]{2016A&A...595A..81M}. The advective nonlinear term, which is also shown in Fig.~\ref{Asinsp7}, is mostly acting around the edges of the plume. Comparing the plots for $|\mathbf{z}_\perp^+|$ and nonlinear advection, we can appreciate the intricate connection between linear processes such as resonant absorption and nonlinear deformations. Advancing the simulation further, we again show snapshots in Fig.~\ref{Asinz30} and Fig.~\ref{Asinsp30}, at $t \approx 1.5 t_A$
In these figures we can appreciate a growing nonlinear behaviour and creation of small scales. In Fig.~\ref{Asinz30}, while the appearance of the surrounding pure Alfv\'en waves is not undergoing changes, $|\mathbf{z}_\perp^+|$ and $|\mathbf{z}_\perp^-|$ at the location of the plume appear increasingly shredded and presenting more intricate structure as the simulation progresses. All the observations made to the earlier snapshots can still be recognized, however there are some important additions. The plume undergoes stronger deformations, obvious in the density plot in Fig.~\ref{Asinsp30}, showing ripple-like structures. Note that while previously $|\mathbf{z}_\perp^-|$ and the nonlinear advection were strongly localized around the edge of the plume, they increasingly become more space-filling, at least inside the plume. \par
\begin{figure}[h]
    \centering     
        \includegraphics[width=1.0\textwidth]{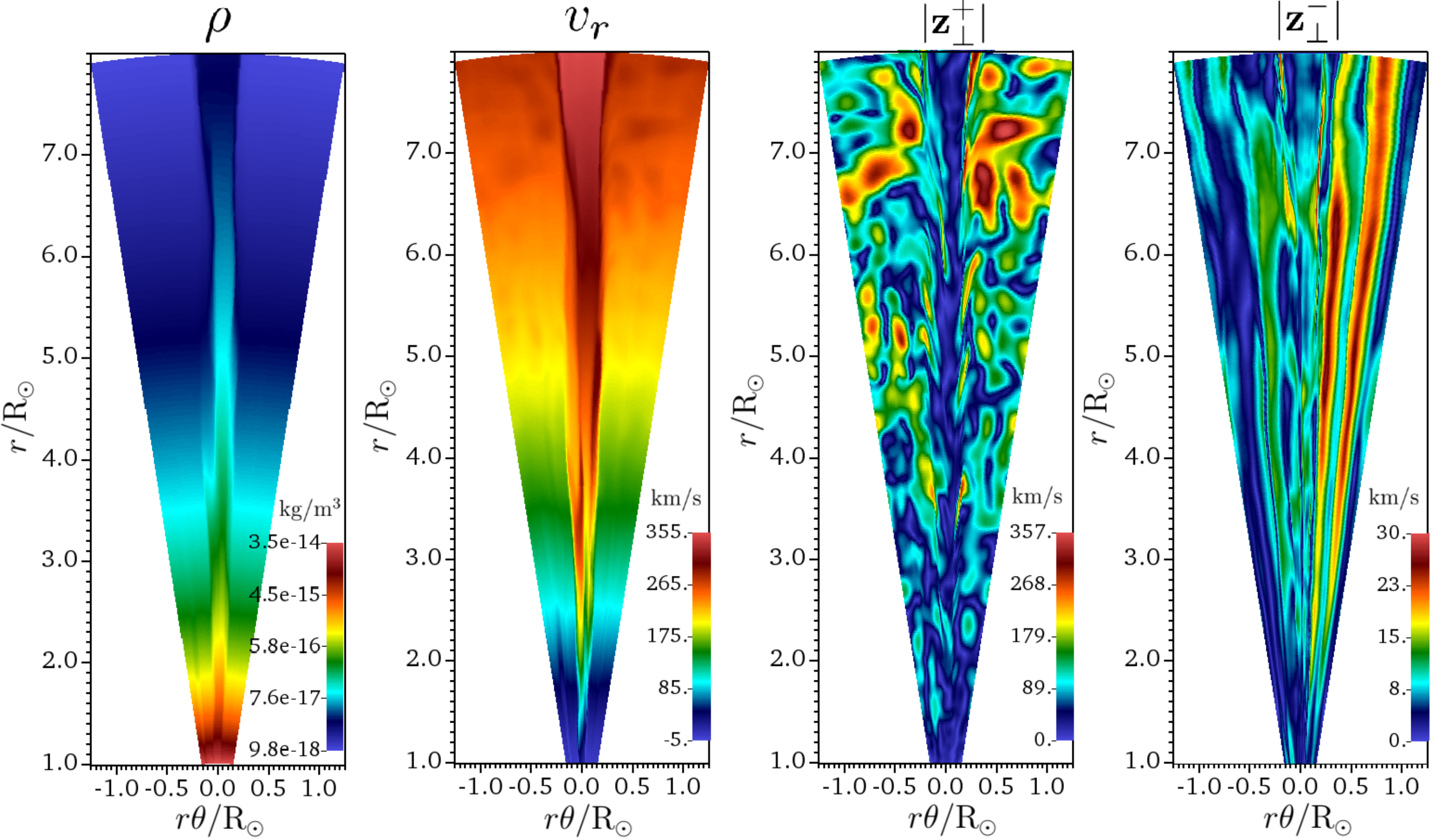}  
        \caption{Snapshots of density ($\rho$), radial velocity ($v_r$), and the perpendicular Els\"{a}sser variables $|\mathbf{z}^+_\perp|$ and $|\mathbf{z}^-_\perp|$, respectively (from left to right), in a slice at $\phi = \pi/2$, $1\ t_A$ after the start of the simulation, in the stochastically-driven plume simulation.}
        \label{Astirz30}
\end{figure}

\begin{figure}[h]
    \centering     
        \includegraphics[width=1.0\textwidth]{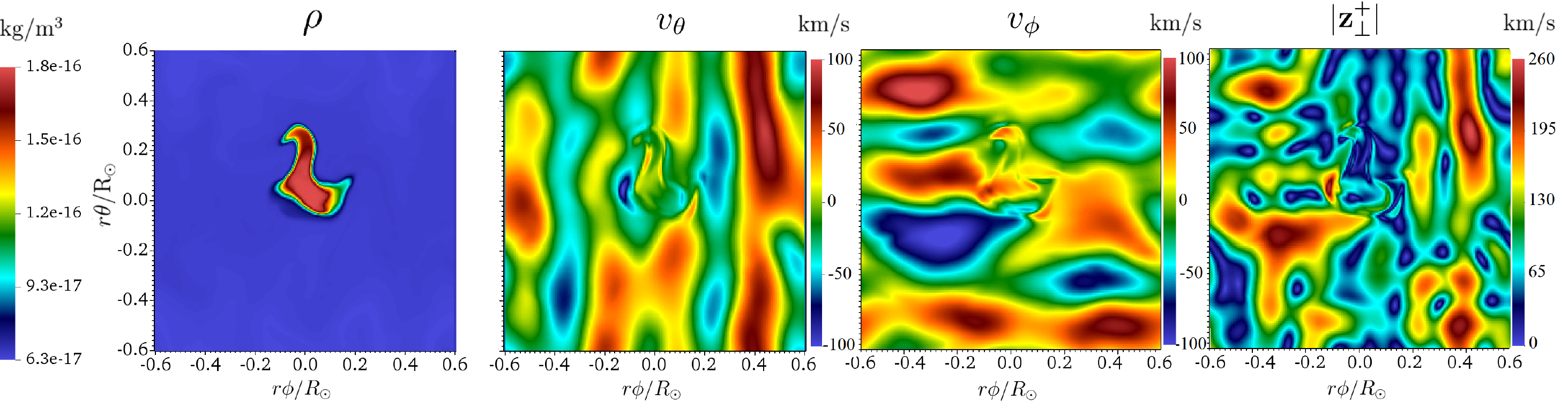}  
        \caption{Snapshots of density ($\rho$), perpendicular velocity components ($v_\theta$ and $v_\phi$), and the perpendicular Els\"{a}sser variable $|\mathbf{z}^+_\perp|$, respectively (from left to right), in a spherical slice at $r = 4\ \mathrm{R}_\odot$, $1.5\ t_A$ after the start of the simulation, in the stochastically-driven plume simulation.}
        \label{Astirsp30}
\end{figure}

\begin{figure}[h]
    \centering     
        \includegraphics[width=0.8\textwidth]{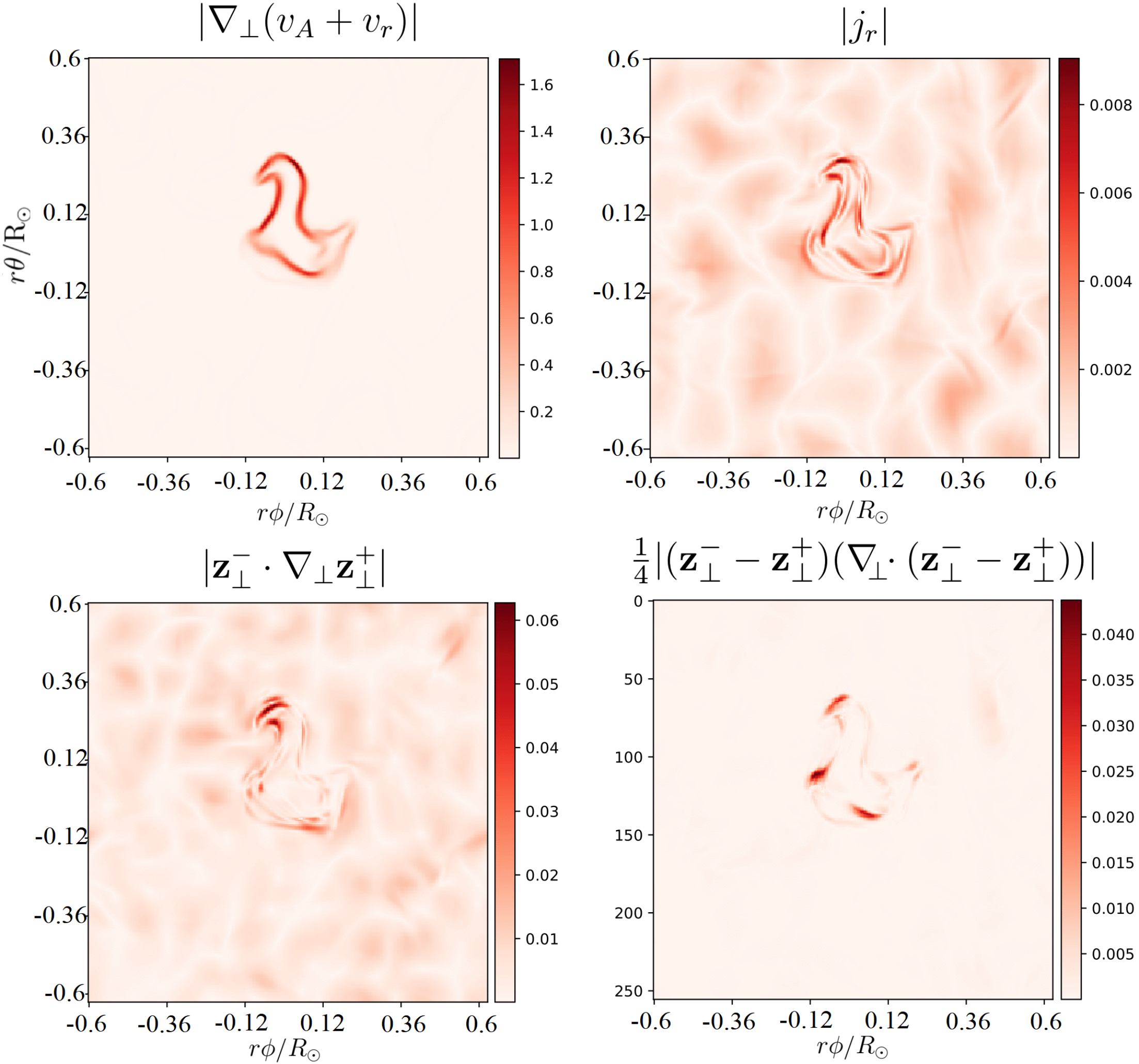}  
        \caption{Snapshots of Doppler-shifted Alfv\'en speed perpendicular gradient (\textit{Top Left}), radial electrical current (\textit{Top Right}), perpendicular nonlinear advection of $\mathbf{z}^+$ (\textit{Bottom Left}), and a nonlinear term related to density inhomogeneities (\textit{Bottom Right}) in spherical slices at $r = 4\ \mathrm{R}_\odot$, at $t\approx 1.5 t_A$. Units are code units.}
        \label{nonlin_stir}
\end{figure}

 
The single-plume simulation with the $k_\perp=0$ sinusoidal driver  shows clearly that $|\mathbf{z}_\perp^-|$ is greatly enhanced by the presence of kink waves, and that most of the nonlinear advection is associated with it. However, although using a simple driver helped to easily identify the different mechansims at work, the nonlinear cascade (by the counterpropagating wave scenario) of the surrounding Alfv\'en waves was eliminated. This is because $k_\perp=0$ or slab Alfv\'en waves cannot nonlinearly interact with their reflections. In Section~\ref{three}, we have shown that the self-cascade of kink waves is proceeding on faster timescales than the nonlinear advection of pure Alfv\'en waves and their reflections, based on dimensional analysis and also indirectly by comparing the evolution in the inhomogeneous and homogeneous setups. In order to verify these claims, we present the results of a single-plume simulation, using the same stochastic driver as in Section~\ref{two}. The dynamics after $t \approx t_A$ are shown in several snapshots in Fig.~\ref{Astirz30} and Fig.~\ref{Astirsp30}.
Although a kink wave is induced in the plume, it is harder to notice it in Fig.~\ref{Astirz30} than previously as the driver is not linearly polarized and sinusoidal. Nevertheless, we can still observe a strong deformation of the plume, deviating greatly from the initial cylindrical cross-section, as seen in Fig.~\ref{Astirsp30}. Note that, as in the previous simulation with the sinusoidal driver, the kink mode of the plume is not preferentially driven. Ponderomotive perturbations in the radial velocity can still be noticed in Fig.~\ref{Astirz30}. 
The appearance of $|\mathbf{z}_\perp^+|$ and $|\mathbf{z}_\perp^-|$ in Fig.~\ref{Astirz30} are reminiscent of the evolution in Fig.~\ref{sliceZ} of Section~\ref{three}. For example, the strong $k_\parallel \approx 0$ power in $|\mathbf{z}_\perp^-|$, mostly induced by the driver, dominates its appearance. 
Nevertheless, it can still be noticed that there are more small-scale dynamics near the edge of the plume. This is clearer in Fig.~\ref{Astirsp30}, and especially in the perpendicular components of the velocity perturbation, while in the $|\mathbf{z}_\perp^+|$ plot it is much harder to spot. In Fig.~\ref{nonlin_stir}, some nonlinearly relevant quantities are shown.    
As hinted at in Section~\ref{three} with cross-correlation analysis, there appears to be a strong correlation between Alfv\'en speed gradients, radial electric currents, and nonlinear terms. While the regions surrounding the plume are not free of currents or nonlinear advection, these quantities are the strongest and more fine-structured around the plume. In the bottom-right panel of Fig.~\ref{nonlin_stir} we show another nonlinear term, enabled by density variations, as discussed previously \citep{2019ApJ...882...50M}. These plots support the indirect evidence by comparison of inhomogeneous and homogeneous simulations, presented in Section~\ref{three}, that the self-cascade of kink waves, or uniturbulence is dominating the turbulence generation.  
 
\begin{acknowledgements}N.M. was supported by a Newton International Fellowship of the Royal Society. V.M.N. was supported by STFC grant ST/T000252/1. \end{acknowledgements}

\bibliographystyle{apj} % style aa.bst
\bibliography{../Biblio}{} % \begin{tiny}

\end{document}